\documentclass[english,prl,aps,twocolumn,10pt]{revtex4-2}

\usepackage{amsmath}
\usepackage{amssymb}
\usepackage{graphicx}
\usepackage[english]{babel}
\usepackage[colorlinks=true, pdfstartview=FitV, linkcolor=blue, citecolor=blue, urlcolor=blue]{hyperref} 

\newcommand{\ket}[1]{\ensuremath{\left|#1\right\rangle}}
\newcommand{\op}[2]{\ensuremath{\left|#1\right\rangle\!\left\langle#2\right|}}

\newcommand{\ee}[1]{\cdot10^{#1}}
\newcommand{\kB}{k_{\rm B}}

\newcommand{\cI}{NV-1}
\newcommand{\eII}{NV-2}
\newcommand{\gV}{NV-3}
\newcommand{\eIV}{NV-4}
\newcommand{\tip}{NV-5}

\newcommand{\dperp}{\delta_\perp}
\newcommand{\Ex}{\text{E}_\text{x}}
\newcommand{\Ey}{\text{E}_\text{y}}
\newcommand{\ms}[1]{$m_{\mathrm{S}} = #1$}
\newcommand{\x}{\text{x}}
\newcommand{\y}{\text{y}}

\def\C        {{$^{12}$C \/}}
\def\N        {{$^{15}$N$^{+}$ \/}}

\usepackage{array}
\usepackage{verbatim}
\usepackage{oplotsymbl} 
\usepackage{multirow} 
\usepackage{siunitx} 
\usepackage{physics}
\usepackage{chemformula} 

\def\Ctwelve    {{$^{12}$C \/}}

\def\second   {{2$^{\rm nd}$ \/}}

\def\etal     {{\it et al.}}

\newcommand{\unitformat}[1]{\,\mathrm{#1}}

\newcommand{\degree}{^\circ}

\newcommand{\kcps}{\text{kcps}}

\begin{document}

\title{Temperature dependence of photoluminescence intensity and spin contrast in nitrogen-vacancy centers}

\author{S.~Ernst$^{1,\dagger}$, P.~J.~Scheidegger$^{1,\dagger}$, S.~Diesch$^{1}$, L.~Lorenzelli$^{1}$, and C.~L.~Degen$^{1,2}$}
\email{degenc@ethz.ch}
\thanks{$^\dagger$These authors contributed equally.}
\affiliation{$^1$Department of Physics, ETH Zurich, Otto Stern Weg 1, 8093 Zurich, Switzerland.}
\affiliation{$^2$Quantum Center, ETH Zurich, 8093 Zurich, Switzerland.}

\begin{abstract}
We report on measurements of the photoluminescence (PL) properties of single nitrogen-vacancy (NV) centers in diamond at temperatures between 4--300\,K.  We observe a strong reduction of the PL intensity and spin contrast between ca. 10--100\,K that recovers to high levels below and above.  Further, we find a rich dependence on magnetic bias field and crystal strain.
We develop a comprehensive model based on spin mixing and orbital hopping in the electronic excited state that quantitatively explains the observations.  Beyond a more complete understanding of the excited-state dynamics, our work provides a novel approach for probing electron-phonon interactions and a predictive tool for optimizing experimental conditions for quantum applications.
\end{abstract}

\date{\today}

\maketitle


The long coherence time~\cite{herbschleb2019} and the ease of optical spin readout have made the negatively charged nitrogen-vacancy (NV) center in diamond a preferred qubit for applications in quantum metrology~\cite{schirhagl14} and quantum information~\cite{childress13}. Extraordinarily, the NV retains its quantum properties up to above room temperature, suggesting its use in both ambient and cryogenic environments.  At room temperature, researchers have employed the NV's spin as a sensor for magnetic~\cite{maze08,balasubramanian08} and electric fields~\cite{dolde11}, and thermometry~\cite{kucsko13, acosta10}. Cooled to below $10\unitformat{K}$, spin-dependent optical transitions~\cite{robledo11nature} have facilitated the implementation of prototypical quantum networks~\cite{pompili21} and multi-qubit quantum operations~\cite{bradley19}. Additionally, cryogenic NV magnetometry has been performed at the micron-~\cite{lillie20}  and nanoscale~\cite{song21, thiel19}.

While the photodynamics of NV centers at low temperature (below $10\unitformat{K}$) and around room temperature have been studied in detail, the understanding in the intermediate temperature range is incomplete.  Initial studies of the photoluminescence (PL) emission intensity of NV ensembles revealed a minimum around $25\unitformat{K}$ attributed to time-averaging in the electronic excited state (ES)~\cite{rogers09}. This averaging process is caused by phonon-mediated transitions between the two orbital branches~\cite{fu09, abtew11}.
A temperature-dependent reduction in PL intensity and spin contrast was also reported in connection with NV charge state instabilities~\cite{wise21}.  Further, spin mixing in the ES due to magnetic field~\cite{tetienne12, happacher22} or crystal strain~\cite{tamarat08} was identified as another mechanism for loss of PL.  The strain-related spin mixing at low temperature was found to be partially mitigated by application of a large magnetic bias field~\cite{happacher22,vool21,rogers09}.
Because high PL intensity and spin contrast are essential for high-fidelity quantum readout and sensitive magnetometry, a complete picture of the NV photodynamics in the $10-100\unitformat{K}$ range is highly desirable.


In this Letter, we report measurements of the PL intensity and spin contrast for single NV centers between $4-300\unitformat{K}$.  We show that a combination of orbital hopping and spin mixing in the ES leads to a strong reduction of both quantities between $10-100\unitformat{K}$.  Based on measurements at varying magnetic field ($0-200\unitformat{mT}$) and intrinsic strain (ES splitting $2 \times (9-80)\unitformat{GHz}$), we develop a comprehensive theoretical model for the temperature-dependent dynamics of the ES.  Details on the model and simulation framework are given in a companion paper \cite{ernst23modeling}.  
As a result, we are able to quantitatively describe the NV's PL intensity and spin contrast over the complete parameter range of temperature, magnetic field and strain, and find excellent agreement with experimental data.


In our study, we investigate single NV centers situated in nanostructured pillars, which serve to enhance the photon collection efficiency.  Our samples include an array of pillars on isotopically pure \C diamond (\cI{} to \eIV{}, ElementSix) and a scanning tip fabricated from natural-abundance material (\tip{}, \mbox{QZabre}). NV centers are formed by shallow \N ion implantation ($7\unitformat{keV}$) followed by high-temperature annealing ($<\,10^{-8} \unitformat{mbar}$, $880 \unitformat{^\circ C}$, $2\unitformat{h}$). Samples are measured in a dry dilution refrigerator (Setup A) at temperatures between $4-100\unitformat{K}$; an additional study down to $0.35\unitformat{K}$ did not show further variation in the NV behavior~\cite{scheidegger22}.  A second dry cryostat (Setup B) with a temperature range of $30-300\unitformat{K}$ is used to validate the aforementioned measurements and extend the range to room temperature~\cite{lorenzelli21}.  Both setups operate in vacuum without addition of exchange gas ($p<5.5\ee{-5}\unitformat{mbar}$).  Magnetic bias fields, when specified, are applied along the NV symmetry axis.

\begin{figure}
    \includegraphics{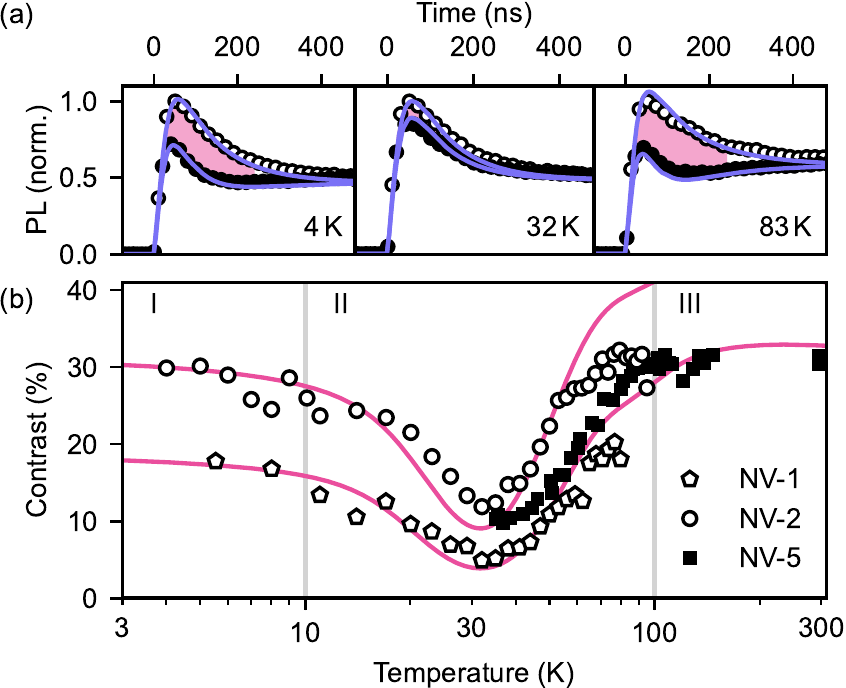}
    \caption{
    (a) Time-dependent PL traces during a laser pulse, measured on \eII{} initialized to the $\ket{0}$ (open circles) and $\ket{-1}$ (filled circles) states at low, intermediate and high temperature.
    The spin contrast is given by the relative difference between the two curves (pink shading).  Solid lines are fits to the PL dynamics.
    (b) Spin contrast versus temperature for three NV centers measured on Setup A (empty markers) and Setup B (filled marker).  Solid lines show corresponding simulations for \cI{} and \eII{}.
    A bias field of $3\unitformat{mT}$ is applied.
		}
    \label{fig1}
\end{figure}

The central experimental observation of this work is reported in Fig.~\ref{fig1}, which plots the spin contrast as a function of temperature $T=4-300\unitformat{K}$.  The PL intensity follows a similar trend (see Fig.~\ref{fig:SI-2Dmaps}), but is more prone to experimental drift.  We measure the contrast by integrating the relative difference in PL between the \ms{0} and \ms{-1} states (subsequently denoted by $\ket{0}$ and $\ket{-1}$) during the first $250\unitformat{ns}$ under excitation with a $520\unitformat{nm}$ diode laser (Fig.~\ref{fig1}(a)).  To initialize the spin state into $\ket{0}$, we use a $ 2\unitformat{\mu s}$ laser pulse, followed by a state swap to $\ket{-1}$ (when needed) using an adiabatic inversion microwave pulse~\cite{kupce95}.
Fig.~\ref{fig1}(b) clearly reveals three temperature regimes:
(I) Below ca. $ 10\unitformat{K}$, the spin contrast is mostly constant.
(II) Between ca. $10-100\unitformat{K}$, the spin contrast is strongly reduced with a pronounced minimum around $35\unitformat{K}$ and then recovers for higher temperatures.
(III) Above ca. $ 100\unitformat{K}$, the spin contrast remains approximately constant up to room temperature.
In all measurements, the room temperature contrast exceeds the low temperature limit.
At even higher temperatures, the contrast has been shown to slowly decrease until it vanishes around $700\unitformat{K}$~\cite{toyli12}.

\begin{figure}
    \includegraphics{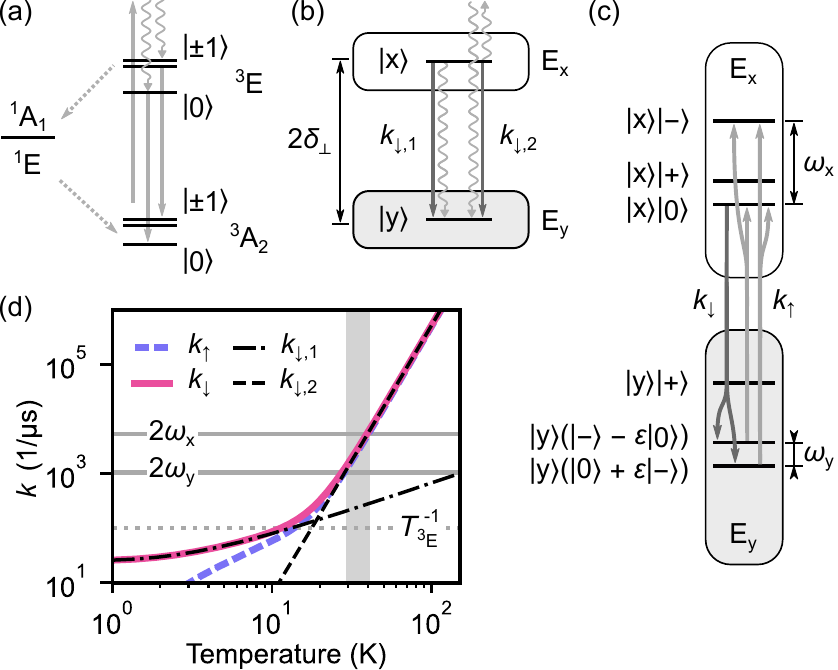}
    \caption{
		(a) Level diagram in the NV spin subspace $ \mathcal{H}_{\text{spin}} $ of the electronic ground ($^3\!$A$_2$) and excited ($^3$E) states, as well as the metastable ($^1\!$A$_1$, $^1$E) shelving states.  The intersystem crossing (dotted) is spin selective, favoring decay out of $\ket{\pm1}$.
		In (a-c), solid arrows mark spin-conserving transitions and curly arrows symbolize phonons.
		(b) Level diagram in the orbital subspace $ \mathcal{H}_{\text{orbit}} $ of the NV ES ($^3$E).  Two orbital branches ($\Ex$, $\Ey$) split under strain $\delta_\perp$.  One- and two-phonon processes cause hopping between branches at temperature-dependent rates $k_{\downarrow,1(2)}$ ($k_{\uparrow,1(2)}$ not shown).
		(c) Example of levels in the composite Hilbert space of orbit and spin $ \mathcal{H}_{\text{orbit}} \otimes \mathcal{H}_{\text{spin}}$. Eigenstates are superpositions of $\ket{0}$ and $\ket{\pm} \propto (\ket{+1}\pm\ket{-1})$.  Spin-conserving, phonon-mediated transitions involving $\ket{0}$ are depicted by gray arrows.  $\omega_\mathrm{x(y)}$ are the Larmor frequencies of involved spin transitions.
		(d) Hopping rates as a function of temperature. Inverse optical lifetime $T_{\mathrm{^3E}}^{-1}$ and Larmor frequencies (in MHz) are indicated by horizontal lines.
		For (c, d) we use $\delta_\perp = 40\unitformat{GHz}$ in the direction of a carbon bond at low magnetic field, such that only two basis states mix significantly ($\left|\varepsilon\right|^2$ = $\left|\epsilon_{\ket{\text{y}, 0}, \ket{\text{y}, -}}\right|^2 = 0.1$ in Tab.~\ref{tab:epsilonij}).
    }
    \label{fig2}
\end{figure}

Before providing a theoretical explanation for the behavior seen in Fig.~\ref{fig1}, we briefly recall the mechanism for contrast generation by looking at the spin-subspace of the NV given in Fig~\ref{fig2}(a)~\cite{doherty13}.  After spin-conserving optical excitation from the ground state (GS) into the ES, a spin-selective intersystem crossing (ISC) leads to preferential population of the shelving state for $\ket{\pm1}$.  Because the shelving state $^1$E has a relatively long lifetime, the average PL emission is lower for $\ket{\pm1}$ compared to $\ket{0}$, leading to spin contrast.  The PL reduction is temporary and disappears due to optical re-pumping into $\ket{0}$ after a few hundred nanoseconds, see Fig~\ref{fig1}(a).  Crucially, this mechanism of contrast generation is effective only for as long as there are no transitions between the ES spin states.

We next consider the orbital subspace of the NV ES ($^3$E), which is a doublet shown in Fig.~\ref{fig2}(b)~\cite{doherty13}. 
In the presence of in-plane strain $\dperp$ relative to the NV principal axis, the ES possesses two orbital branches, $\Ex$ and $\Ey$, split by $2 \dperp$~\cite{batalov09}.  In the composite space of orbit and spin, each branch has three spin states, leading to a total of six energy eigenlevels (Fig.~\ref{fig2}(c)).  We now show that the contrast reduction and recovery can be explained by the interplay of two mechanisms: spin mixing and orbital branch hopping in the ES.

First, we discuss the effects of spin mixing, meaning that the ES eigenstates are not pure spin eigenstates.  As an example, we consider Fig.~\ref{fig2}(c).  Here, the $\ket{0}$ state is in good approximation an eigenstate of the $\Ex$ branch but not the $\Ey$ branch, where it forms a superposition with the $\ket{-} \propto (\ket{+1}-\ket{-1})$ state. Consequently, optical excitation into the $\Ex$ branch is spin-conserving, while excitation into the $\Ey$ branch will lead to spin mixing.  In general, the spin mixing amplitudes $\epsilon_{\ket{i},\ket{j}}$ between basis states $\ket{i}$ and $\ket{j}$ in the six eigenlevels depend on the strain magnitude and direction~\cite{tamarat08, manson06}, as well as magnetic field alignment~\cite{tetienne12} and magnitude~\cite{happacher22}.  Therefore, the spin contrast is both strain and field-dependent.  Although the $\epsilon_{\ket{i},\ket{j}}$ are typically small, they play a key role in the mechanism of spin relaxation.

Second, we consider the effects of orbital hopping, which refers to spin-conserving transitions between $\Ex$ and $\Ey$ driven by phonons.  Fig.~\ref{fig2}(b) schematically depicts the dominant contributions arising from one-phonon processes (rates $k_1$) and two-phonon processes (rates $k_2$) derived in Ref.~\cite{ernst23modeling, goldman15, plakhotnik15}.
The one-phonon downward ($\Ex\rightarrow\Ey$) hopping rate is given by
\begin{equation}\label{eq:kone}
  k_{\downarrow,1}(T, \delta_\perp) \propto \eta \delta_\perp^3 \left[ n(2 h \delta_\perp/\kB T) + 1\right],
\end{equation}
where $\eta$ parametrizes the electron-phonon coupling and $ n $ is the Bose-Einstein distribution function.  The rate of the two-phonon process is given by
\begin{equation}\label{eq:ktwo}
	k_{\downarrow,2}(T) \propto \eta^2 T^5 I(T).
\end{equation}
where $I(T)$ is a mildly strain- and temperature-dependent integral over the phonon spectrum that we solve in the Debye approximation.  The total hopping rates are the sums of the one- and two-phonon contributions, $k_{\downarrow(\uparrow)} = k_{\downarrow(\uparrow),1} + k_{\downarrow(\uparrow),2}$.  The upward ($\Ey\rightarrow\Ex$) rate significantly differs from the downward rate only below $10\unitformat{K}$, where it is reduced by the absence of spontaneous emission (second term in Eq.~\ref{eq:kone}).
Fig.~\ref{fig2}(d) plots the hopping rates for the parameters in Fig.~\ref{fig2}(c) as a function of temperature.  It provides the key to explaining our experimental observations in the temperature regimes (I-III):

(I) Below ca. $10\unitformat{K}$, the orbital hopping is dominated by $k_{\downarrow,1}$ due to the spontaneous emission. Since $k_{\downarrow,1}$ is slower than the ES decay rate ($T_\mathrm{^3E}^{-1} \approx 10^8\unitformat{s^{-1}}$) for typical strain values $\delta_\perp \lesssim 40\unitformat{GHz}$, the ES spin states are mostly preserved (except for some small spin mixing $\epsilon_i $) and the spin contrast is high.

(II) Above $10\unitformat{K}$, the two-phonon process starts to dominate.  Once $k > T_\mathrm{^3E}^{-1}$, spin relaxation between $\ket{0}$ and $\ket{\pm1}$ is drastically amplified, because the time evolution under different Larmor precession in both branches becomes randomized by the frequent hopping.  
This relaxation mechanism is most efficient when a hopping event occurs approximately every half of a Larmor precession period $(2\omega_\mathrm{x(y)})^{-1}$~\cite{ernst23modeling}.
This occurs between ca. $30-40\unitformat{K}$ (gray shading in Fig.~\ref{fig2}(d)) and coincides with the temperature where we observe the strongest suppression of the spin contrast.

(III) As the temperature increases further, the orbital hopping rates become much faster than the spin dynamics and the two orbital states are time-averaged~\cite{rogers09, batalov09}.  This effectively renders $^3$E an orbital singlet similar to the GS $^3\!$A$_2$~\cite{plakhotnik14} and leads to the commonly accepted room-temperature model appearing as in Fig.~\ref{fig2}(a).  Since $\ket{0}$ and $\ket{\pm1}$ are pure eigenstates of the time-averaged Hamiltonian, the highest spin contrast is observed in this regime. 

Armed with this theory, we implement a rate model to quantitatively reproduce the experimental observations by numerical simulations.  Details on the rate model and simulations are given in a companion paper \cite{ernst23modeling} and the Supplemental Information.  We model the orbital hopping by spin-conserving Markovian transitions between the two orbital branches.  Since spin coherences are maintained during the transitions, we use a Lindblad master equation rather than a classical rate model. We describe the ES in a composite Hilbert space of spin and orbit ($ \mathcal{H}_\mathrm{ES} = \mathcal{H}_{\text{orbit}} \otimes \mathcal{H}_{\text{spin}}$) and formulate the spin-conserving jump operators as
\begin{align}
L_{\downarrow}^\text{ES} &= \sqrt{k_{\downarrow, 1}+k_{\downarrow, 2}} \op{\y}{\x} \otimes \mathbb{I}_3 \ , \label{equ:LmixDown}
\end{align}
and likewise for $L_{\uparrow}^\text{ES}$.  We further introduce optical excitation, decay and ISC by classical jump operators. The resulting Liouville equation describes the time evolution of the 10-dimensional density matrix $\rho(t)$, containing three GS levels, six ES levels and one combined shelving state.

\begin{figure*}
    \includegraphics{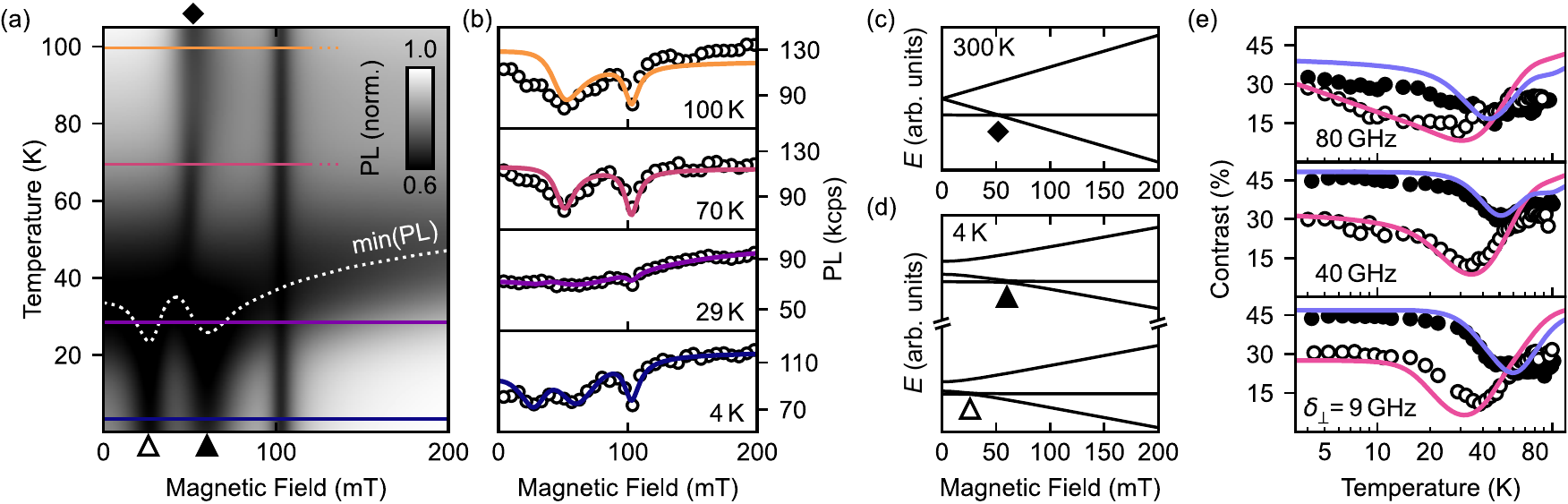}
    \caption{
		(a) Simulation of the PL in dependence of the magnetic field and temperature. The PL is strongly reduced at avoided crossings of the excited state (symbols) and the ground state ($103\unitformat{mT}$) energy levels. The white dotted line marks the PL minimum, coloured solid lines indicate the line cuts shown in (b). The simulation is based on parameters fitted to \cI{}. 
    (b) Experimental PL curves for \cI{} measured as a function of magnetic field and temperature. Solid lines are fits.
    (c,d) Energy levels for the \cI{} ES at $300\unitformat{K}$ (c, time-averaged) and at $4\unitformat{K}$ (d).  Symbols refer to (a).
		(e) Experimental spin contrast as a function of temperature for NV centers with different strain $\dperp$.
		Measurements are taken at $3\unitformat{mT}$ (empty circles) and $200\unitformat{mT}$ (filled circles).
		Solid lines show the corresponding simulations.
		}
    \label{fig3}
\end{figure*}

To simulate the behavior of a chosen NV center, we feed our model with values obtained from a simultaneous fit of three sets of calibration measurements:
(i) We use a measurement of the steady-state PL intensity as a function of magnetic field at base temperature (see Fig.~\ref{fig3}(b, $4\unitformat{K}$)) to obtain strain values and unintended misalignment of the bias field, fitting the minima in the PL at level anti-crossings~\cite{rogers09, happacher22}.
(ii) We pick a set of 24 time-dependent PL traces (c.f. Fig.~\ref{fig1}(a)), including two spin states ($\ket{0}$, $\ket{-1}$), six temperatures ($4-100\unitformat{K}$), and low and high bias field ($3\unitformat{mT}$, $200\unitformat{mT}$).  Fits to these PL traces then yield the optical decay and ISC rates, which are approximately temperature-independent~\cite{goldman15prb} and are known to vary between NV centers~\cite{doherty13, robledo11njp}, as well as the coupling strength $\eta$.
We determine the shelving state lifetime~\cite{robledo11njp, manson06}, which has a mild, well-known temperature dependence, in a separate calibration.
(iii) For each time-dependent PL trace, we perform an optical saturation measurement to quantify drift in the
background luminescence, optical alignment, and ratio of collection over excitation efficiency.
Finally, we use literature values for the NV fine structure~\cite{bassett14, chen11apl}.

As an important side result, our calibration yields values for the electron-phonon couplings $\eta$ ranging from $176\unitformat{\mu s}^{-1}\unitformat{meV}^{-3}$ (\eII{}, used in Fig.~\ref{fig2}(d)) to $268\unitformat{\mu s}^{-1}\unitformat{meV}^{-3}$ (\eIV{}), in good agreement with Refs.~\cite{abtew11, plakhotnik15, goldman15}.  We note that these studies use different phonon models in the evaluation of $I(T)$.  While our data does not allow validation of a particular model with certainty, our measurement approach provides complementary insight into $I(T)$~\cite{ernst23modeling} (see also Fig.~\ref{fig:SI-cutoff}).

We are now ready to return to Fig.~\ref{fig1}(b) and use our model and calibration to simulate the temperature-dependent PL and spin contrast (solid curves).  Overall, we find an excellent agreement between experimental and simulated results.  In particular, the model quantitatively reproduces all temperature regimes (I-III), including the minimum in contrast around $35\unitformat{K}$ and the recovery towards room temperature.  Although the agreement is not perfect at elevated temperatures, which we attribute to setup instabilities and uncertainty in temperature calibration (see Supplemental Information~\ref{sec:experimental}), our model successfully bridges the classical rate models used in the the limits of low~\cite{happacher22} and high~\cite{tetienne12,robledo11njp} temperatures.


Next, we use our model to predict the PL properties as a function of magnetic bias field.  In Fig.~\ref{fig3}(a,b), we plot the simulated PL intensity as a function of $B=0-200\unitformat{mT}$ and $T=0-100\unitformat{K}$ together with the experimental results. The model successfully predicts the known reductions in PL (indicated by symbols) at magnetic fields that correspond to level-anti-crossings (LAC) in the ES, in both the low (Fig.~\ref{fig3}(d)) and high temperature limit (Fig.~\ref{fig3}(c), obtained from Fig.~\ref{fig3}(d) by a partial trace over the orbital subspace).
The global PL minimum -- indicated by the dotted line in Fig.~\ref{fig3}(a) -- depends on the exact energy level spacing at a given magnetic field (and strain). For example, we observe that with increasing magnetic field (beyond the second ES LAC), the PL minimum becomes less pronounced and shifts to higher temperatures. This behavior is readily explained by a lower degree of spin mixing in the eigenstates (smaller $\epsilon_{\ket{i},\ket{j}}$ in Fig.~\ref{fig2}(c)) and higher Larmor frequencies (larger $\omega_\mathrm{x(y)}$ in Fig.~\ref{fig2}(d)) at high field.  However, even at the highest field accessible in our experiment ($200\unitformat{mT}$), the PL minimum is still noticeable.  Full recovery of the PL is expected for fields significantly above $1\unitformat{T}$ (Fig.~\ref{fig:LowStrainAndHighBMaps}).


Finally, we examine the influence of crystal strain.  In Fig.~\ref{fig3}(e), we compare the temperature dependence of the spin contrast for NV centers with high (\eIV{}, $80\unitformat{GHz}$), medium (\eII{}, $40\unitformat{GHz}$), and low (\gV{}, $9\unitformat{GHz}$) intrinsic strain within our accessible range (\cI{} has $32\unitformat{GHz}$).  While all curves show the same qualitative behavior, we find that the most prominent feature is a decrease in the spin contrast at high strain $\dperp$ already below 10\,K.  This feature can be understood through the factor $ \dperp^3$ in Eq.~\ref{eq:kone}: $k_{\uparrow,1}$ is rapidly increasing as the required high energy phonon modes become thermally activated, approaching $k_{\downarrow,1} $, which is generally high due to spontaneous emission (Fig.~\ref{fig:SI-hopping-rates}).


In conclusion, we developed a rate model that explains the NV center photo-physics over a broad range of temperature, magnetic bias field and crystal strain, and find excellent agreement with the experiment.  In particular, our model successfully predicts a minimum in the PL emission and spin contrast around $35\unitformat{K}$ due to rapid spin relaxation driven by an interplay of spin mixing and orbital hopping.
This spin relaxation process degrades both the spin initialization and the spin readout fidelity (Fig.~\ref{fig:SI-2DmapsInitReadout}).
It is of fundamental nature and universal to all NV centers, including NV centers deep in the bulk that experience negligible crystal strain~\cite{batalov09} (Fig.~\ref{fig:LowStrainAndHighBMaps}).

Our work provides useful insight beyond giving a more complete picture of the NV excited-state dynamics.
Firstly, our model can account for the observed temperature dependence by phonon-induced processes in the ES alone.  Therefore, we conclude that charge-state switching between NV$^-$ and NV$^0$ does not play a key role in explaining the spin contrast as a function of temperature.  We also have not observed any signs of charge state instabilities on the few-minutes time scale of our measurements (see Fig.~\ref{fig:SI-histograms}).  
Second, our work introduces a new measurement approach for probing electron-phonon interactions and contributing modes, applicable in regimes where resonant laser PL excitation spectroscopy~\cite{goldman15} or measurement of motional narrowing on ES ODMR lines~\cite{plakhotnik15} are unavailable.
Third, we examined the rich dependence on magnetic field, strain (or equivalently electric field~\cite{doherty13}), and temperature.  Here, our model offers a predictive tool for maximizing the PL intensity and spin contrast, which are the key quantities for achieving high spin readout fidelity and high metrology sensitivity in quantum applications.

\textit{Note added: } We acknowledge related work on the temperature dependence of the NV photo-physics by Happacher \textit{et al.}~\cite{happacher2023} and Blakley \textit{et al.}~\cite{blakley2023_arxiv}.

The authors thank Matthew Markham (ElementSix) for providing the $^{12}$C diamond, Jan Rhensius (QZabre) for nanofabrication, and Erika Janitz, Fedor Jelezko, Assaf Hamo, Konstantin Herb, William Huxter, Patrick Maletinsky, Francesco Poggiali, Friedemann Reinhard, J\"org Wrachtrup and Jonathan Zopes for useful input and discussions.
This work was supported by the European Research Council through ERC CoG 817720 (IMAGINE), the Swiss National Science Foundation (SNSF) through Project Grant No. 200020\_175600 and through the NCCR QSIT, a National Centre of Competence in Research in Quantum Science and Technology, Grant No. 51NF40-185902, and the Advancing Science and TEchnology thRough dIamond Quantum Sensing (ASTERIQS) program, Grant No. 820394, of the European Commission.

%

\clearpage
\begin{widetext}
\setcounter{figure}{0}
\renewcommand{\thefigure}{S\arabic{figure}}
\renewcommand{\theequation}{S\arabic{equation}}

\vspace*{.3cm}
\begin{center}
\huge
Supplemental Material
\end{center}

\tableofcontents

\newpage
\section{Experimental}
\label{sec:experimental}
The majority of data presented in this work was acquired inside a dilution refrigerator (Setup A) on an isotopically pure (\Ctwelve) diamond sample doped with $7\unitformat{keV}$ NV centers in nanostructured pillars. To minimize heating by the microwave excitation, we employ an impedance-matched co-planar waveguide (CPW) made from thin-film aluminum on top of a sapphire substrate, which is identical to the CPW we used in our recent sub-Kelvin scanning magnetometry experiments~\cite{scheidegger22}. To ensure good thermalization, we glue the diamond sample directly on top of the CPW, which in turn is glued on a sample holder made from copper. The sample holder contains both, a resistive heater and a calibrated thermometer (LakeShore Cernox). A PID controller is used to stabilize the temperature in a range from $4\unitformat{K}$ to $100\unitformat{K}$ with a temperature stability of 0.1K at low temperatures and $< 1$K at very high temperatures. Towards higher measurement temperatures ($> 60 \unitformat{K}$), we increasingly find thermal drifts, which we stabilize by frequent optical tracking. Despite our efforts, one can observe that the contrast values measured in the high temperature limit are systematically lower than the simulated model (Fig.~1(b) and Fig.~3(e) of the main text). We attribute this to increased setup instabilities and possibly some uncertainty in the measured temperature, due to the large temperature gradient between the sample and the rest of the cryostat (Setup A).

For the optical excitation of the NV, we use a $520 \unitformat{nm}$ diode-laser with home-built modulation circuitry. We characterize its rise time on fluorescent contamination in proximity to the NV center and obtain $\tau_R = 23 \pm 1\unitformat{ns}\,.$ We calibrate the laser power directly at its output and use a conversion factor to fit the actual (lower) laser power on the NV. Heating of the diode slightly alters the output laser power, which is why we find a consistent mismatch by a factor of $1.15$ between the steady-state PL value in \textit{time-resolved pulsed ODMR} and the \textit{saturation} measurements (for details, see section \ref{sec:measurements}) and we scale our data accordingly.

We use a superconducting vector magnet (American Magnetics Inc.) to apply a magnetic field along the respective NV axis.
The field is aligned by sweeping both spherical angles at a constant field magnitude of $B=200\unitformat{mT}$ and minimizing the ODMR resonance of the \ms{-1} state . This calibration has been performed once for every NV. 
As common for superconducting magnets, we observe a remanent field of approximately $3\unitformat{mT}$ (projected onto the NV axis) after operation at elevated magnetic fields (here: $200\unitformat{mT}$). We can reset the magnet by heating it to above the superconducting transition temperature and letting it cool down again. In our experiments, we perform such a reset when switching to a different NV (i.e. moving the magnetic field orientation) and when lowering the magnetic field magnitude. For measuring the PL as a function of magnetic field (c.f. \ref{par:plvb}), we sweep from high to low field while maintaining the same orientation (without resetting). 

We note that our data was acquired over the course of three distinct cooldowns. In cooldown \#1, we used a different microwave antenna than the CPW discussed above, which resulted in microwave heating. Consequently, we resorted to all-optical measurements of PL during this cooldown. Cooldowns \#2 and \#3 were identical in terms of experimental setup and used the aluminum CPW. All contrast vs temperature data on Setup A were taken during those cooldowns. During cooldown \#2, two of our NVs (\eII{} and \eIV) bleached (i.e. they showed a complete loss of spin contrast at a reduced PL level). Bringing them to ambient conditions for a short time completely restored their previous properties and they have not shown signs of bleaching since. Importantly, all our data is consistent across all three cooldowns and in line with the observations in Setup B (\tip), which emphasizes the independence of our results from setup-related conditions.

\section{Measurements}
\label{sec:measurements}
In our work, we use five types of experiments to thoroughly characterize each NV and to disentangle the various fitting parameters.
Fitting the first three experiment types (\ref{par:plvb} - \ref{par:saturation}) is a simultaneous effort, further discussed in section \ref{sec:fitting}, due to shared parameters between all of them. The fourth experiment type, namely the the shelving state lifetime (\ref{par:ssl}), is determined independently.
The fifth type of experiment (\ref{par:histograms}) is used to verify that we are working with the negative charge state NV$^-$ of the NV center.
This section discusses all five experiments in detail.

\subsection{PL vs. \textit{B}}
\label{par:plvb}
We measure the steady-state PL while sweeping the magnitude of the magnetic field $B$ from high field to low field (Fig.~\ref{fig:SI-superfit}(a)). At base temperature, such a trace exhibits minima at the level anti-crossings (LAC) of the ES and the GS, which we use to uniquely characterize the in-plane strain and magnetic field alignment~\cite{rogers09, happacher22}.

While the relative depth of the ES LAC minima at base temperature (e.g. $\SI{30}{\milli\tesla}$ and $\SI{55}{\milli\tesla}$ in Fig.~\ref{fig:SI-superfit}(a)) is known to depend on the ES branch-selectivity of the optical excitation~\cite{happacher22}, we find an additional dependence on the orbital hopping rate. Specifically, the spontaneous emission process $ \Ex \rightarrow \Ey $ that is relevant in the presence of in-plane strain causes the first $\Ey$ LAC minimum to be deeper than the second $\Ex$ LAC minimum. 
The rate of this spontaneous emission process (c.f. Eq.~\ref{equ:k1hoppUpDown}) depends on the electron-phonon coupling strength $ \eta $, which is a fit parameter in our model. To avoid cross-talk between the fit of $ \eta $ and the branch selectivity, we always assume no orbital branch selectivity in our optical excitation, i.e. $ r_\beta = \beta_\text{x}/\beta_\text{y} = 1 $ (c.f. Fig.~\ref{fig:SIrates}). This assumption is reasonable for the following reasons: (i) we off-resonantly excite NV centers with green laser light into the phonon sideband, where the selectivity is naturally low even with aligned NV and laser polarization axes~\cite{fu09, ulbricht16}. (ii) our NV center principal axes are tilted by $55\degree$ relative to the diamond surface and optical axis, reducing the possible polarization alignment.

When fitting a \textit{PL vs. $B$} measurement, we exclude points below $15\unitformat{mT}$, to avoid possible distortions due to the remanent field of the vector magnet.

\subsection{Time-resolved pulsed ODMR}
\label{par:trodmr}
We measure the time-resolved, spin state dependent PL under excitation with $ \sim\SI{2}{\micro\second} $ laser pulses (Fig.~\ref{fig:SI-superfit}(b)). After a such a laser pulse, the spin state of the NV center is initialized -- with some laser power dependent fidelity \cite{wirtitsch23} -- into the \ms{0} state.
In a subsequent laser pulse, we thus obtain the time-resolved PL of the \ms{0} initialized state.
Likewise, if a microwave pulse that results in an adiabatic inversion of the spin state is applied beforehand, we obtain the time-resolved PL of the \ms{-1} initialized state.
The relative difference between these two traces, integrated over the first $250\unitformat{ns}$, results in the ODMR spin contrast.
Explicitly, the contrast is calculated from the integrated counts of the \ms{-1} state divided by the integrated counts of the \ms{0} state.
We find that the contrast is fairly resilient against changes in setup-specific parameters.
In addition to determining contrast, we use the raw, time-resolved PL traces to fit for the rates related to the optical lifetime and intersysten-crossing (ISC) process. We note that setup-specific parameters also change the curve shape.

\subsection{Saturation measurement}
\label{par:saturation}
The measurement of the steady-state PL while sweeping the laser power shows a characteristic saturation behavior that arises when the optical excitation rate exceeds the lifetime of the electronic excited state. Then, the laser power dependent PL becomes linear with a slope given by the fluorescent background (c.f. Fig.~\ref{fig:SI-superfit}(c)). The laser power at which the saturation is reached is characteristic for the excitation efficiency. The absolute PL level on the other hand is characteristic for the collection efficiency. Therefore, saturation measurements are primarily suited to characterize these three setup-specific parameters (background, excitation and collection efficiency) and to disentangle them from rate parameters.

\subsection{Shelving state lifetime}
\label{par:ssl}
The shelving state lifetime (SSL), i.e. the time spent in the intermediate singlet levels before decaying back into the GS (c.f. Fig.~2 of the main text), has a well-known temperature dependence that we include in our model of the NV photo-physics. We measure the SSL following Refs.~\cite{robledo11njp, manson06}, by first exciting the NV into the shelving state using a $1.2\unitformat{\mu s}$ laser pulse and subsequently probing the initial PL in a second laser pulse, delayed by $\tau$. The resulting exponential rise $\mathrm{PL} \propto 1 - a e^{-\tau/\tau_S(T)}$ can be fitted for the SSL $\tau_S(T)$ at the measurement temperature $T$.
Obtaining such data points $\tau_S(T)$ for several temperatures allows in a second step to fit the parameters of the temperature dependence of the SSL.
We fit $ \tau_{S,0} $ at zero temperature from Eq.~\ref{equ:SSL} for each NV center individually. The fit results are given in table \ref{tab:fitedparams}. Fitting $ \tau_S (T) $ reliably for $ \Delta E $ in Eq.~\ref{equ:SSL} requires a significant amount of measurements in the range $ T \gg 100\unitformat{K} $, which was not accessible in our experiments. We therefore use the same $ \Delta E = 16.6 \unitformat{meV} $ as reported by \citet{robledo11njp}.
In Fig.~\ref{fig:SI-superfit}(d) we present the result of such an evaluation scheme at the example of \eII.

\begin{figure*}
    \includegraphics{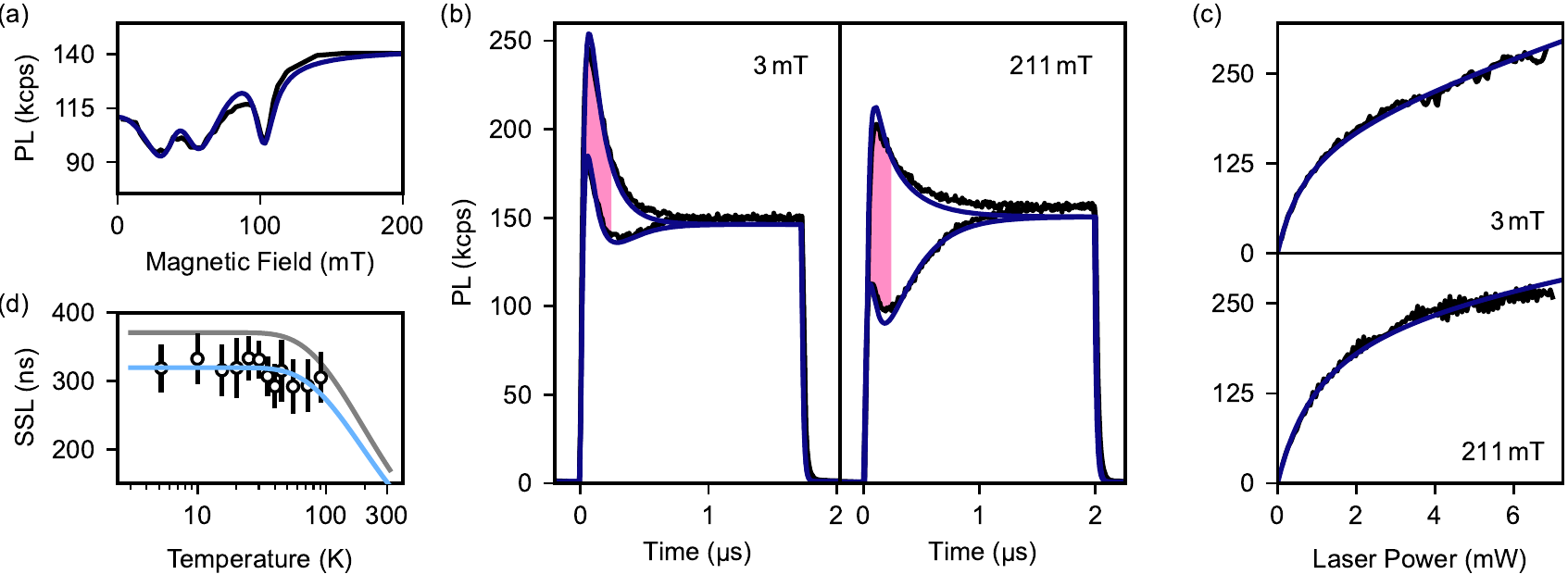}
    \caption{\textbf{Example for measurement types used in our fitting process} - Shown  are complementary measurements used in this work, with their corresponding fits (solid blue lines), taken on \eII{}. (a~-~c) were taken at base temperature ($\sim 5\unitformat{K}$) and constitute a typical data set for the simultaneous fitting described in section \ref{sec:fitting}. (a) \textit{PL vs. $B$} measurement. (b) \textit{Time-resolved pulsed ODMR} at low and high magnetic field. Contrast is integrated over the pink shaded area (first $250\unitformat{ns}$). (c) \textit{Saturation} measurement. (d) Measurement and fit of the temperature dependence of the SSL. For comparison, the findings of \citet{robledo11njp} are also shown (grey).
		}
    \label{fig:SI-superfit}
\end{figure*}

\subsection{Continuous time tagging}
\label{par:histograms}
At each temperature step, after all other measurements are completed, we additionally record PL during $60\unitformat{s}$ of continuous laser illumination. The raw timetags (resolution: $10 \unitformat{ns}$) are binned with a sampling rate of $10\unitformat{kHz}$ to obtain count-rate histograms as shown in Fig.~\ref{fig:SI-histograms} (for \cI{} at select temperatures). After removing noise from pulse tube vibrations ($1.4\unitformat{Hz}$) and rotary valve motor vibrations ($140\unitformat{Hz}$) and their harmonics, the histograms approximate the shot noise limit well. The strictly Gaussian nature of the signal indicates that no blinking occurred on our NVs on time scales limited by the $60\unitformat{s}$-long measurement time on slow time-scales, and the $10\unitformat{kHz}$ sampling rate on fast time-scales. This supports our conclusion in the main text, that charge-state switching between NV$^-$ and NV$^0$ is not involved in the temperature-dependent reduction of PL and contrast.

\begin{figure*}
    \includegraphics{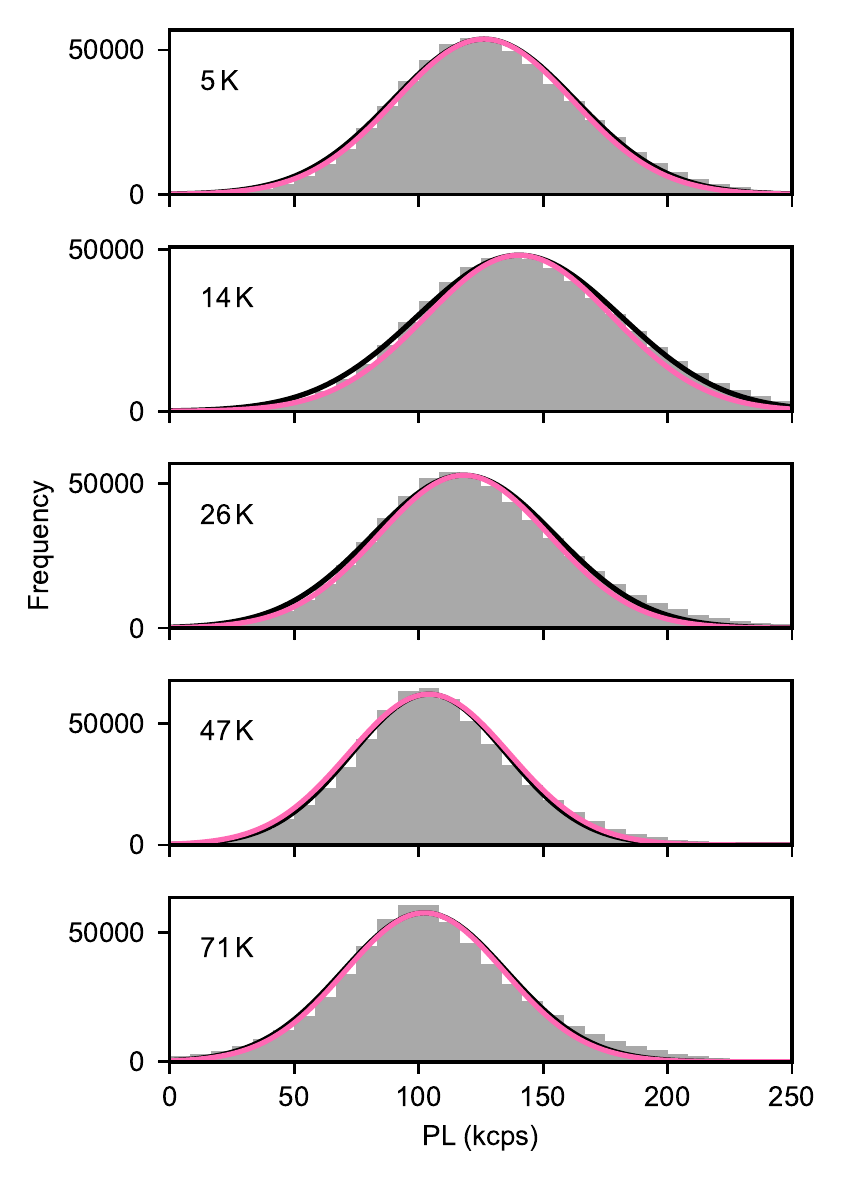}
    \caption{\textbf{Histograms of the PL count rate of \cI{} at varying temperatures} - Dilution refrigerator vibrations at $1.4\unitformat{Hz}$ and $140\unitformat{Hz}$ and their harmonics have been removed from the data. The pink curve indicates the theoretical shot noise limit of the measurement, the black line is a Gaussian fit to the histogram.} 
    \label{fig:SI-histograms}
\end{figure*}

\section{The numerical model}
Our model is derived in detail in Ref.~\cite{ernst23modeling}. Here, we summarize the most important equations and relate them to our fitting algorithm.

\subsection{Level structure and optical transition rates}
We calculate the NVs energy level structure and eigenstates at a given magnetic bias field (with magnitude $B$, polar angle $\theta_B$ and azimuthal angle $\phi_B$) and in-plane strain (with magnitude $\delta_\perp$ and azimuthal angle $\phi_\delta$) using the well-established low temperature Hamiltonian~\cite{doherty13}. We neglect the hyperfine interactions and on-axis strain, as they are not relevant for this work. Angles are measured relative to the NV coordinate system in Ref.~\cite{doherty13}. Between the calculated levels, we define transition rates to model optical excitation ($\beta_\mathrm{x/y} k_r$) and relaxation ($k_r$, $k_{E_{12}}$, $k_{E_{\mathrm{xy}}}$, $k_{A_1}$), as well as phonon-mediated spin-conserving transitions ($k_{\uparrow/\downarrow}$), as depicted in Fig.~\ref{fig:SIrates}. To do so, it is convenient to implement the Hamiltonian in different bases and transform between them.

We assume no temperature dependence for the optical excitation $k_r$~\cite{blakley2023_arxiv}. The intersystem-crossing (ISC) transitions from the excited state to the shelving state (SS) have been shown to be slightly temperature dependent \cite{goldman15, goldman_erratum_2017} at high temperature, but the effect is too small to have an impact on our model. The temperature dependence of the ISC rates below $\SI{30}{\kelvin}$, as reported in their work, is readily contained in our model~\cite{ernst23modeling}.

The SS relaxation, on the other hand, has a well established temperature dependence and we model it following \citet{robledo11njp}  
\begin{equation}\label{equ:SSL}
	\tau_S (T) = \frac{1}{k_{S0} + k_{S1}}  =  \tau_{S,0} \left( 1 - e^{-\frac{\Delta E}{\kB T}} \right)
\end{equation}
with a spontaneous emission process ($ \tau_{S,0} $) and a stimulated emission process of a phonon with energy $ \Delta E $. Here, $k_\mathrm{B}$ is the Boltzmann constant and we use the literature values for $\Delta E$ (see table \ref{tab:fitedparams}). We note that the branching ratio $r_S = k_{S0}/k_{S1}$ remains approximately constant in temperature \cite{gali19}.

\begin{figure}
	\centering
	\includegraphics[width=0.7\linewidth]{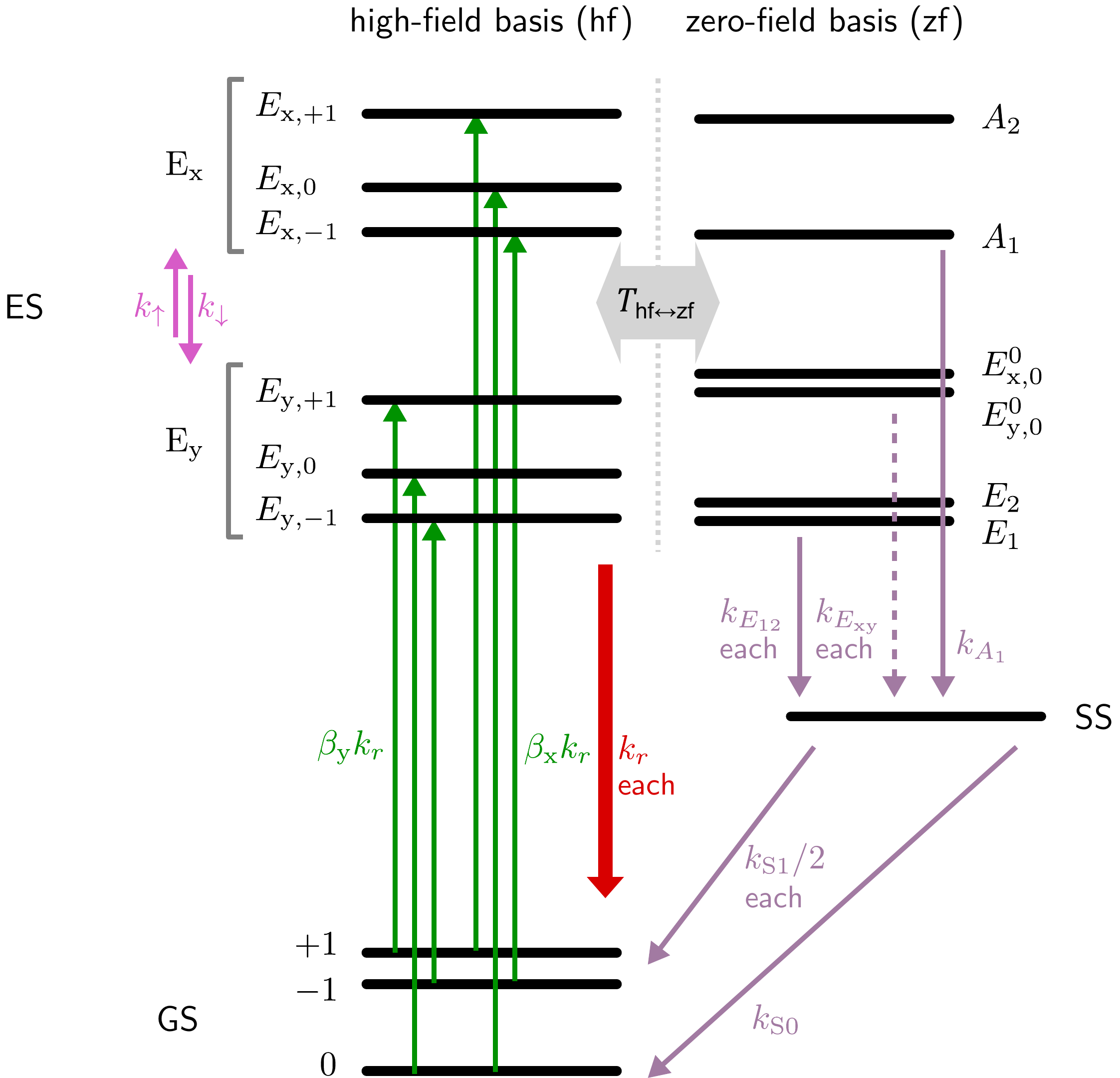}
	\caption{\textbf{Energy levels and transition rates} -
			Depicted are the 10 energy levels of the zero-temperature \ch{NV-} center Hamiltonian. The ES is depicted in two different bases: the high-field basis (hf) which is a good eigenbasis at high magnetic field $ \vec{B} \parallel \hat{e}_z $ and high strain $ \delta_\perp $ and the zero-field basis (zf), which covers the limit of $ \vec{B} \rightarrow 0 $ and $ \delta_\perp \rightarrow 0 $. Transition rates (arrows) are implemented in the basis in which they are defined and we use basis transformations to combine them into a single master equation (c.f. section~\ref{sec:ME}).
            Optical excitation ($ \beta_{\text{x/y}} k_r$) from the ground state GS to the two branches ($\Ex$, $\Ey$) of the excited state ES is spin $ m_s $ conserving and dependent on the excitation power and polarization (x, y). Optical decay ($k_r$) is also spin conserving and leads to the emission of a red photon. The ISC rates ($k_{E_{12}}$, $k_{E_{\mathrm{xy}}}$, $k_{A_1}$) are dependent on the fine structure levels, which are good eigenstates in zero-field. The decay from the SS to the GS repopulates all spin states, with a branching ratio defined as $r_S = k_0/k_{S1}$.
            Phonon mediated, coherent spin state preserving transitions ($k_{\uparrow/\downarrow}$) between the two orbital branches occur at a rate that increases with temperature.
	}
	\label{fig:SIrates}
\end{figure}

\subsection{Orbital hopping rates}
The primary effect of temperature in our model lies in the addition of hopping rates between the two orbital branches to the rate model, similar to previous studies~\cite{fu09, goldman15, plakhotnik15}. This hopping is caused by one- and two-phonon-processes and thus depends on the thermal occupation of phonon modes. 
We introduce rates	
\begin{equation}\label{equ:khoppUpDown}
    \begin{split}
	\Ey \rightarrow \Ex &: k_{\uparrow}(T, \delta_\perp, \eta) = k_{\uparrow,1} + k_{\uparrow,2} \\
	\Ex \rightarrow \Ey &: k_{\downarrow}(T, \delta_\perp, \eta) = k_{\downarrow,1} + k_{\downarrow,2}
    \end{split}
\end{equation}
which depend on both temperature $ T $ and the strain induced splitting of the orbital branches, which is approximately $\approx 2\delta_\perp$ for a sufficiently small $g_l B_z$, as we expect for all our measurements~\cite{happacher22}. The coupling strength between electronic orbital states of the \ch{NV-} and phonons is given by $ \eta $.

The expressions for the hopping rates can be derived following Fermi's Golden Rule, as is done in Refs.~\cite{ernst23modeling, goldman15prb, plakhotnik15}. Here, we just state the results. For the one-phonon process, one finds
\begin{equation}\label{equ:k1hoppUpDown}
    \begin{split}
	k_{\uparrow,1}(T, \delta_\perp) &\approx 32 \eta h^3 \delta_\perp^3  n(2 \delta_\perp h) \\
	k_{\downarrow,1}(T, \delta_\perp) &\approx 32 \eta h^3 \delta_\perp^3 \left[ n(2 \delta_\perp h) + 1\right]\,,
    \end{split}
\end{equation}
where the Bose–Einstein distribution $n(\epsilon)$ describes the thermal occupation of phonon modes with energy $ \epsilon $ at temperature $T$. The rates for the opposite directions are related by detailed balance as $k_{\uparrow} = \text{exp}\left(-\frac{ 2\delta_\perp h}{\kB T} \right) k_{\downarrow}$. Notably, the spontaneous emission term (``$ +1 $'') in $k_{\downarrow,1}$ results in a finite, strain-dependent rate that needs to be considered even at $ T=\SI{0}{\kelvin} $.

For the two-phonon process, we find
\begin{equation}\label{equ:k2hoppUp}
\begin{aligned}
	k_{\uparrow,2}(T,\delta_\perp) = &\frac{64\hbar}{\pi} \eta^2 \kB^5 T^5 \int_{x_\perp}^{\Omega/\kB T} \frac{e^{x}x(x-x_\perp)\left[x^2+(x-x_\perp)^2\right]}{2 \left(e^{x}-1\right) \left(e^{x-x_\perp}-1\right)} dx\\
	= & \frac{64\hbar}{\pi} \eta^2 \kB^5 T^5 I(T, \delta_\perp)\,,
\end{aligned}\\
\end{equation}
where $x$ is the phonon energy in units of $k_\text{B} T$ and $x_\perp \approx \frac{2\delta_\perp h}{\kB T}$.
In the second line, we introduced the temperature- and strain-dependent integral over the phonon modes $I(T, \delta_\perp)$. Using the detailed balance, one can directly obtain $ k_{\downarrow,2}(T,\delta_\perp) $.

\subsection{Master equation}
\label{sec:ME}
Importantly, the orbital hopping rates $ k_{\uparrow/\downarrow} $ only act on the orbital subspace, leaving the state in the spin subspace untouched. This cannot be modeled by a classical rate equation model as is commonly used. We therefore employ a master equation model that acts on quantum states but can include the known classical rates via jump operators. We numerically calculate the time evolution of the 10-dimensional density matrix $\rho$ (including the three GS states, six ES states and the SS) using the Liouville equation
\begin{equation}\label{equ:ME}
	\dv{t} \hat{\rho} = - \frac{i}{\hbar} \comm{\hat{H}}{\hat{\rho}} + \sum_{k} \left( \hat{L}_k \hat{\rho} \hat{L}_k^\dagger - \frac{1}{2} \acomm{\hat{L}_k^\dagger \hat{L}_k}{\hat{\rho}} \right) \equiv \mathcal{L}(\rho)\,.
\end{equation}
Details on the implementation of the jump operators can be found in Ref.~\cite{ernst23modeling}.

\subsection{Linking simulation with experiment}
\label{sec:Linking}
The quantity we observe in our experiments is always the $ \text{PL} $ averaged over some time or many repetitions of a sequence of laser and microwave pulses. The PL is composed of radiative emission of the \ch{NV-} center and fluorescent background radiation.
We assume a linear dependence of the background $ \text{PL} $ on the laser power. 
The collected $ \text{PL} $ can be calculated from the populations $ \rho_{i,i} $ of the emitting levels, which have the same indices $ i \in [4, 9] $ for all bases. We thus model the observed PL as:
\begin{equation}\label{equ:PL}
	\text{PL}(t) = A R \left(\sum_{i \in [4, 9]} \rho_{i,i} (t) k_r \right) + b P
\end{equation}
where we introduce a set of setup-related parameters. $A$ represents the optical excitation efficiency (unit: $\text{W}^{-1}$), which can drift over time and change from experiment to experiment if setup conditions are not stable enough.
$R$ is the ratio of collection over excitation efficiency (unit: cps W s) and is  expected to be fairly stable against drifts in the optical path. Lastly, $b$ characterizes the background counts (unit: $\text{cps W}$), which are linear in the laser power $P$.

To simulate and fit the time-resolved PL response of NV centers in our setup as given in Fig.~1(a) of the main text, we need to take the rise time of our laser pulses into account.
Thus, we introduce the laser rise time $ \tau_R $ to our model, the value of which we determined experimentally. This is necessary when fitting $ \text{PL}(t) $ time traces as $ \tau_R $ of our laser is longer than the time resolution of our photo-detector ($ 10\unitformat{ns} $).
In the derivation of our model we assume a time-independent Liouvillian superoperator $ \mathcal{L} $ in the Liouville equation~\ref{equ:ME}.
The laser power $ P $ enters into $ \mathcal{L} $ in the form of a pre-factor of the jump operators describing the optical excitation. We thus need to ramp up the laser power and thus $ \mathcal{L} $ in $ N $ discrete steps
\begin{equation}\label{equ:tauR}
	P(t) =
	\begin{cases}
		&0 \text{ if } t<t_0 \\
		&P \left( 1 - e^{\frac{t_n}{\tau_R}} \right) \text{ for } t_n = \Bigl\lceil \frac{t - t_0}{\Delta t} \Bigr\rceil \Delta t  \\
		&P \text{ if } t>t_0+N\tau_R
	\end{cases}
\end{equation}
and use fix parameters $ N=4 $ and $ \Delta t=5\unitformat{ns} $ here.
We find that the influence of $ \tau_R $ on the spin contrast is negligibly small. In fitting time-resolved pulsed ODMR traces, on the other hand, $ \tau_R $ is crucial. Likewise, we need to determine the precise value of $t_0$ individually for each measurement. This is done by a linear extrapolation to $ \text{PL}=0 $ from the first two data points above the dark count level, which we also determine and include in the model.

\subsection{Evaluating sensing performance}
\label{sec:sens}
In this work, we aim to give a direct comparison of the performance of the \ch{NV-} center as a sensor over a vast range of parameters. To that end, it is necessary to compare the optimized performance for each parameter setting, as done in section \ref{sec:sensitivity}. We optimize the integration time of simulated pulsed ODMR experiments, which are representative for all kinds of sensing schemes. We note though, that also the laser power could be optimized, which is beyond the scope of this work.
In the following, we give a brief derivation of the sensitivity and its relation to the SNR.

We assume a pulsed ODMR experiment to measure the magnetic field component along the NV axis~\cite{dreau11}.
The sensitivity $s = \Delta B_{\text{min}} \sqrt{t}$ is then given by the minimal magnetic field change $\Delta B_{\text{min}}$, that can be measured within time $t$.
The microwave frequency $f$ dependent photon counts signal $\mathcal{S}$ has the shape of a Gaussian with
\begin{equation*}
    \mathcal{S}(f) = \left[1-e^{-4 \ln(2) (f-f_0)^2/\nu^2} \right] \mathcal{S}_\text{\ms{0}}\,,
\end{equation*}
where $f_0$ is the resonance frequency at which a pi-pulse between the spin levels can happen, and $\nu$ is the full width half maximum (FWHM) of the Gaussian (and related to the inverse spin coherence time). The $\mathcal{S}$ ($\mathcal{S}_\text{\ms{0}}$) are the total collected counts during $t$ that fall into the integration window $t_\text{int}$ with (without) a microwave pulse applied before the laser pulse (c.f. Fig.~\ref{fig:SI-superfit}(b)). Since $f_0$ shifts with the magnetic field as $\Delta B = \Delta f\, 2 \pi/\gamma  = \Delta f/\SI{28.025}{\mega\hertz\per\milli\tesla}$, best (i.e. smallest) sensitivity is achieved at the frequency $f_s$ of highest slope of $\mathcal{S}(f)$. At this frequency, a pulsed ODMR experiment is performed, repetitively consisting of a pi pulse and a subsequent laser pulse for readout and re-initialization. Assuming a shot noise $\Delta \mathcal{S} = \sqrt{\mathcal{S}}$ limited readout at $f_s$, one finds for the sensitivity
\begin{equation}\label{equ:sens}
    s = \frac{2 \pi}{\gamma} m_G^{-1} \frac{\nu \sqrt{r_G}}{C \sqrt{\text{PL}_\text{\ms{0}}} \sqrt{\text{DC}}}\,.
\end{equation}
Due to the Gaussian shape, $m_G^{-1} = 0.700$ and $ r_G= 1- 0.607 \cdot C$. Further, $\text{PL}_\text{\ms{0}} = \mathcal{S}_\text{\ms{0}}/(t_\text{int} N_\text{seq})$ is the average PL count rate during $t_\text{int}$, with $ N_\text{seq} = t/T_\text{seq}$ the number of repetitions of the pulsed ODMR sequence of duration $T_\text{seq}$. Finally, the readout duty cycle $\text{DC} = t_\text{int}/T_\text{seq}$ and the spin contrast $C=1-\mathcal{S}(f_0)/\mathcal{S}_\text{\ms{0}}$.

Viewing the readout of the spin state in a generalized and ideal scheme~\cite{hopper18} a similar result of 
\begin{equation}\label{equ:sensIdeal}
    s \propto \text{SNR}^{-1} \propto \frac{\sqrt{r}}{C \sqrt{\text{PL}_\text{\ms{0}}}}
\end{equation}
can be found for the sensitivity and SNR. Here, $r=1- 0.5 \cdot  C$.

In simulations with optimal readout, we optimize Eq.~\ref{equ:sens} at every parameter setting for the $t_\text{int}$ that gives best sensitivity. Explicitly, this yields not the $t_\text{int}$ with highest contrast.
For a more intuitive picture, the inverse sensitivity and SNR are approximately proportional to
\begin{equation}\label{equ:sensApprox}
    \text{SNR} \sim C \sqrt{\text{PL}_\text{ss}}\,,
\end{equation}
with the quantities spin contrast $C$ and steady-state $\text{PL}$ (here called $\text{PL}_\text{ss}$ for clarity) used elsewhere in this work.

\section{Data fitting and simulation}
In this section, we describe how we fit our measurements to calibrate the model and then, subsequently, simulate the NV PL to compare the theory to our measurements. When fitting, we distinguish two primary sets of fit parameters. The first set, the \textit{robust} parameters, includes all NV-intrinsic parameters (i.e. strain and rates) and the magnetic bias field, which we have good control over. These parameters are stable against scanning stage movements and other drifts which can occur between measurements, particularly when changing to a new temperature or magnetic field. The second set, the \textit{environment-sensitive} parameters, is prone to change between measurements. They include the setup parameters discussed in section \ref{sec:Linking}. These definitions help to distinguish the following three processes:

\begin{itemize}
    \item \textit{Calibration}: Simultaneous fit of multiple data sets across multiple temperatures $T_i$ and fields $B_i$ with common \textit{robust} parameters  but unique \textit{environment-sensitive} parameters for each $T_i$/$B_i$. The procedure is described in great detail in the next section \ref{sec:fitting}.
    \item \textit{Simulation}: Uses the calibration-results for the \textit{robust} parameters and a \textit{single} set of \textit{environment-sensitive} parameters to effectively extrapolate the NV photo-physics to the full range of temperature, magnetic field and strain. This is used, for example, in contrast vs. temperature curves (c.f. Fig.~1(b) in the main text) and in maps of PL vs. $B$ vs. $T$ (c.f. Fig.~3(a)). Deviations from experimental data are found, particularly at high temperatures where drifts become significant.
    \item \textit{Fits}: Fits use the calibration-results for the \textit{robust} parameters but exclusively re-fit the \textit{environment-sensitive} parameters to a given data set at a specific ($T$, $B$). This effectively corrects for experimental drifts.  Examples are fits of \textit{PL vs. $B$} (c.f. Fig.~3(b)) measurements and PL traces (c.f. Fig.~1(a)).
\end{itemize}

\subsection{Calibration}
\label{sec:fitting}

The NV center Hamiltonian that we use in our model contains eight fundamental constants describing the various interactions that yield its spin and electronic eigenstates at zero temperature. We use recent literature values for these~\cite{doherty13, bassett14, chen11apl, rogers09} and assume them to be the same for all conditions. Apart from those constants, there are several other parameters in the model:  (i) NV center specific crystal strain parameters, (ii) magnetic bias field related parameters, (iii) rates, (iv) electron-phonon interaction, and finally (v) setup specific parameters. A subset of those parameters is fixed during the complete calibration process described below, either because they are \textit{predetermined} from separate measurements (e.g. laser rise time) or because they are taken from literature. An overview of all parameters (fitted and fixed) and their respective categorization (robust, environment-sensitive, predetermined, literature) is given in Tab.~\ref{tab:fitedparams}.

We perform the fitting in three stages (I-III), where each stage informs the set of initial parameters of the subsequent fit. This ensures good convergence despite the vast number of parameters.

(I) We start by fitting a \textit{PL vs. $B$} measurement at base temperature ($\sim 5\si{K}$) to our full model.
We fit for the in-plane strain ($\delta_\perp$, $\phi_\delta$) and magnetic field alignment ($\theta_B$, $\phi_B$) as well as the setup-specific parameters ($b$, $R$, $A$). We use common literature values for all other parameters. 

(II) We then simultaneously fit a data set consisting of five different measurements at base temperature, with all parameters being fitted (except for $\eta$). This data set includes (i) the \textit{PL vs. $B$} from before, (ii) two \textit{time-resolved pulsed ODMR} traces and (iii) two \textit{saturation} measurements at low ($3\unitformat{mT}$) and at high ($200\unitformat{mT}$) magnetic field. Such a data set for \eII{} is shown in Fig.~\ref{fig:SI-superfit}.

(III) Finally, we fit the same base-temperature \textit{PL vs. $B$} together with 6 (5 for \eIV{}) pairs of \textit{time-resolved pulsed ODMR} and \textit{saturation} measurements, sampled across our full temperature range. This allows us to fit for the electron-phonon coupling strength $\eta$, in addition to a final fine-tuning of the full parameter set. Crucially, we use a common set of \textit{robust} parameters across all data sets, but individual \textit{environment-sensitive} parameters are used for each data set, which account for mechanical drifts in the experimental setup. Specifically, the two parameters background $b$ and collection over excitation ratio $R$ are fitted separately to each pair of \textit{saturation} and \textit{time-resolved pulsed ODMR} measurements (at the same field and temperature) and also separately to the base-temperature \textit{PL vs. $B$} measurement. Further, the alignment parameter $A$ has to be fitted separately for every single measurement. This results in a total of 60 fit parameters simultaneously fitted to 25 distinct measurements. 

We find that the simultaneous fit of these 25 measurements in stage (III) improves the stability of the fit compared to stage (II) containing only the five base temperature measurements. Particularly, the in-plane magnetic field angle, which results from finite misalignment, can have little observable effect on the PL at base temperature but plays a more significant role at higher temperature.

\subsection{Calibration results}
\label{sec:fitRes}

\begin{table}[t]
\centering
\sffamily
\begin{tabular}{l l r r r r r c}
	& & \textbf{\cI{}} & \textbf{\eII{}} &  \textbf{\gV{}} & \textbf{\eIV{}} \\
	\multicolumn{6}{l}{\textbf{Strain}} \\ [0.5ex]
	$ \dperp $ & ES in-plane strain $ (\si{\giga\hertz}) $ & $ \SI{31.8}{} $ & $ \SI{39.9}{} $ & $ \SI{8.7}{} $ & $ \SI{80.2}{} $ & & \textit{a} \\
	\hline		
	$ \phi_\delta $ & ES in-plane strain angle & $ \SI{39.9}{\degree} $ & $ \SI{24.4}{\degree} $ & $ \SI{0.8}{\degree} $ & $ \SI{59.9}{\degree} $ & & \textit{a} \\ [1.5ex]
	
	\multicolumn{6}{l}{\textbf{Magnetic bias field}} \\ [0.5ex]
	$ \theta_B $ & mag. field alignment angle & $ \SI{1.7}{\degree} $ & $ \SI{1.9}{\degree} $ & $ \SI{1.8}{\degree} $ & $ \SI{2.2}{\degree} $ & & \textit{a} \\
	\hline
	$ \phi_B $ & mag. field in-plane angle & $ \SI{244.0}{\degree} $ & $ \SI{17.6}{\degree} $ & $ \SI{106.5}{\degree} $ & $ \SI{130.6}{\degree} $ && \textit{a} \\
	\hline
	$ \phi_B $ & cooldown \#1 & $ \SI{256.2}{\degree} $ & $ \SI{194.2}{\degree} $ &  && & \textit{a}  \\ [1.5ex]
	
	\multicolumn{6}{l}{\textbf{Rates}} \\ [0.5ex]
	$ k_r $ & opt. emission rate $ (\si{\per\micro\second}) $ & $ \SI{55.1}{} $ & $ \SI{55.7}{} $ & $ \SI{45.3}{} $ & $ \SI{40.5}{} $ && \textit{a} \\
	\hline
	$ k_{E_{12}} $ & ISC rate from $E_{1, 2}$ $ (\si{\per\micro\second}) $ & $ \SI{112.4}{} $ & $ \SI{98.7}{} $ & $ \SI{101.3}{} $ & $ \SI{90.7}{} $ && \textit{a} \\
	\hline
    $ k_{A1} $ & ISC rate from $A_1$ $ (\si{\per\micro\second}) $ & \multicolumn{4}{c}{$k_{E_{12}}/0.52$ \cite{goldman15}} && \textit{d} \\
	\hline
	$ k_{E_{\text{xy}}} $ & ISC rate from $ E_{\mathrm{x, y}} $ $ (\si{\per\micro\second}) $ & $ \SI{9.1}{} $ & $ \SI{8.2}{} $ & $ \SI{8.6}{} $ & $ \SI{7.5}{} $  && \textit{a} \\
    \hline
	$ r_\beta =  \beta_\text{x}/\beta_\text{y}$ & opt. exc. branching ratio & \multicolumn{4}{c}{$1$} && \textit{c} \\
	\hline
	$ r_S = k_{S0}/k_{S1}$ & SS branching ratio & $ \num{1.36} $ & $ \num{2.26} $ & $ \num{1.44} $ & $ \num{1.15} $ && \textit{a} \\
	\hline
	$ \tau_{S,0} = 1/(k_{S0} + k_{S1}) $ & SS decay time at $ T = \SI{0}{\kelvin}$ $ (\si{\nano\second}) $  & $ 342 $ & $ 320 $ & $ 292 $ & $ 318 $ && \textit{c} \\
    \hline
    $ \Delta E $ & SS emit. phonon energy (meV) & \multicolumn{4}{c}{16.6 \cite{robledo11njp}} 
 && \textit{d} \\[1.5ex]
	
	\multicolumn{6}{l}{\textbf{Electron-phonon interaction}} \\ [0.5ex]
	$ \eta $ & ES coupling strength $(\si{\per\micro\second\per\cubic\milli\electronvolt})$ & $ \SI{197}{} $ & $ \SI{176}{} $ & $ \SI{268}{} $ & $ \SI{249}{} $ && \textit{a} \\
    \hline
    $ \Omega $ & phon. cutoff energy (meV) & \multicolumn{4}{c}{168} && \textit{d}\\[1.5ex]

	\multicolumn{6}{l}{\textbf{Setup} - Base temperature PL vs. $B$ measurement} \\ [0.5ex]
	$ b $ & background ($ \kcps\,\si{\per\milli\watt} $) & $ \num{40.8} $ & $ \num{27.5} $ & $ \num{52.6} $ & $ \num{34.7} $  && \textit{b} \\
	\hline
	$ R $ & $ \frac{\text{collection}}{\text{excitation}} $ efficiency ($ \kcps\,\si{\milli\watt\micro\second} $) & $ \num{67.1} $ & $ \num{88.4} $ & $ \num{112.1} $ & $ \num{178.3} $ && \textit{b} \\
	\hline
	$ A $ & opt. alignment $ (\si{\per\watt}) $  & $ \SI{245.1}{} $ & $ \SI{136.2}{} $ & $ \SI{282.3}{} $ & $ \SI{188.1}{} $ && \textit{b} \\
    \hline
    $ \tau_R $ & laser rise time (ns) & \multicolumn{4}{c}{23} && \textit{c} \\
\end{tabular}
\caption{\textbf{Calibration results per NV center} - We fit the NV center intrinsic parameters individually for each center since they are known to vary between NV centers~\cite{doherty13}. The fit results for setup related parameters of saturation and time-resolved pulsed ODMR measurements of the same simultaneous fits are presented in Fig.~\ref{fig:SI-setup}. The last column specifies the type of parameter: (a) robust parameters, including NV intrinsic properties and the well-controlled magnetic bias field; (b) environment-sensitive parameters, which include optical excitation and collection efficiencies; (c) pre-determined parameters; (d) literature values (Hamiltonian parameters not shown). The latter two (c, d) are fixed when fitting for (a) and (b).}
\label{tab:fitedparams}
\end{table}

\begin{figure*}
    \includegraphics{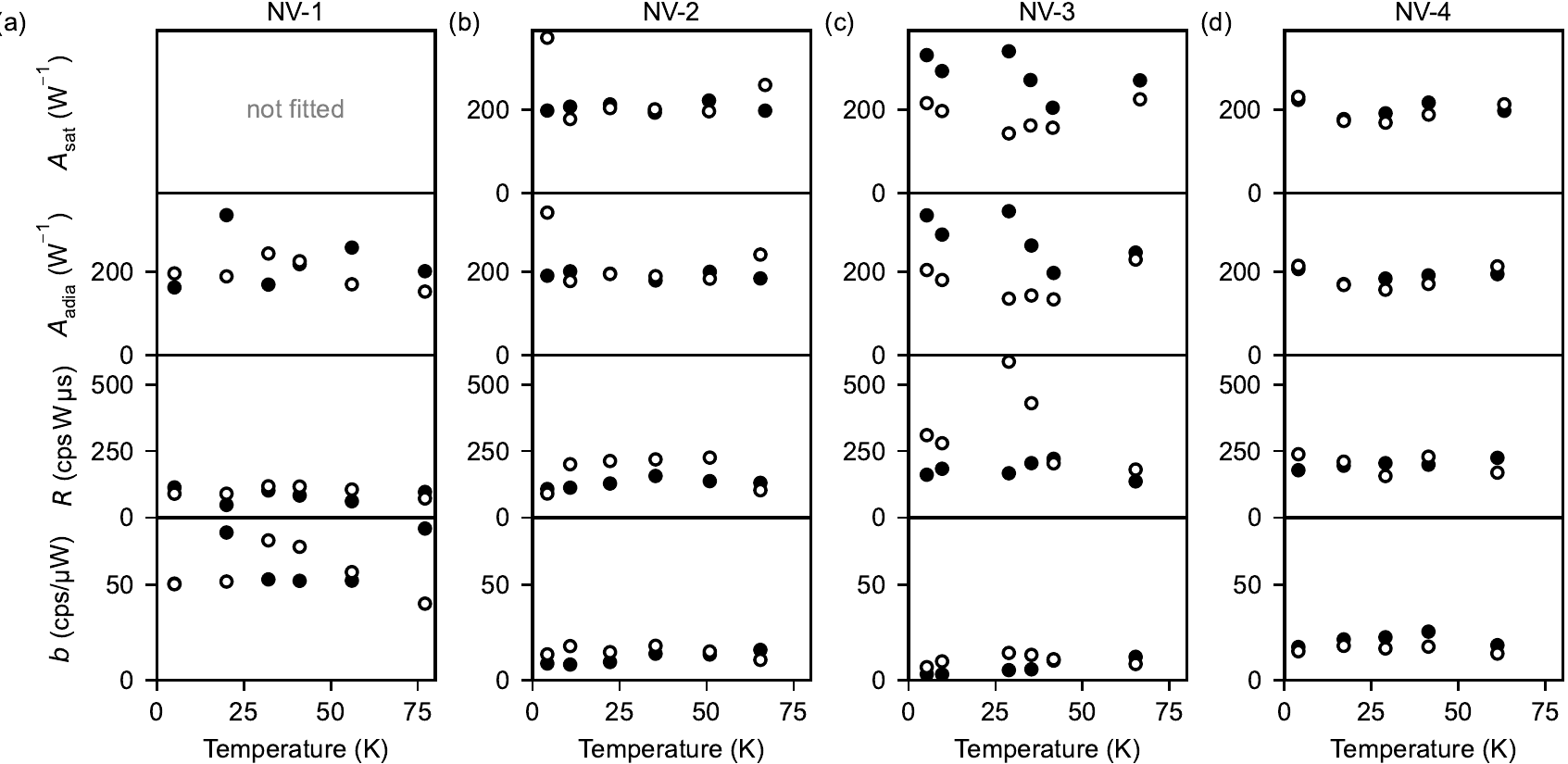}
    \caption{\textbf{Calibration results for the \textit{environment-sensitive} parameters per NV center} - Shown are the results for the \textit{saturation} (``sat'') measurements and \textit{time-resolved pulsed ODMRs} (``adia'') at low ($\SI{3}{\milli\tesla}$, empty ciclres) and at high ($\SI{200}{\milli\tesla}$, filled circles) magnetic field. Temperature-independent results of the same calibration per NV center are given in Tab.~\ref{tab:fitedparams}.
    Shown here are the optical alignment $A$, the ratio of collection over excitation efficiency $R$ (common to both ``sat'' and ``adia'') and background $b$ (also common to both).
		}
    \label{fig:SI-setup}
\end{figure*}

Results of the calibration for the different NV centers are presented in Tab.~\ref{tab:fitedparams} and Fig.~\ref{fig:SI-setup}. The latter contains the \textit{environment-sensitive} parameters that have been determined across all temperatures, as described above.
For \cI{} we were not able to include \textit{saturation} measurements in the fit as the PL stability required here was not sufficient.
We stress that the fitted setup-parameters do not show a correlation with temperature. In particular, in Fig.~\ref{fig:SI-setup} we do not see any feature at $\sim 30\unitformat{K}$, the location of the contrast and PL minimum. 

Since cooldown \#1 has different magnetic field alignment due to changes in the setup, a different in-plane magnetic field angle was used for the \textit{PL vs. $B$} measurement at base temperature if fitted simultaneously with measurements from cooldown \#2 and \#3 -- this is the case for \eII{} and \cI{}. 
Based on our measurements, we can only determine the in-plane magnetic field and strain angle relative to one out of three possible crystal axes, which we then call the $x$-Axis of our coordinate system~\cite{doherty13}. Moreover, we do not know the sign of the $z$-Axis. Therefore, given the three-fold rotational symmetry and mirror symmetry to the $xz$-plane, we choose to determine values for the in-plane strain angle in the range $[0\degree, 60\degree]$. The in-plane magnetic field angle can then have any orientation relative to it, i.e. covers a range of $[0\degree, 360\degree]$.

We compare the calibration results of select parameters to literature values in table \ref{tab:literaturecomp}. 
First, we note that our SS branching ratio $r_S$ fits well into the broad range of values obtained in other experiments. It appears that $r_S$ varies substantially between NV centers even within individual studies. Theoretical modelling prefers a higher $r_S \approx 6$ \cite{gali19} than measured in many of those works, including ours. The cause for this disparity is unknown, and no significant strain or temperature dependence has been observed to date. This represents an open question that will have to be addressed with suitable measurement techniques in future work.

Second, our measured electron-phonon coupling strength $\eta$ falls right into the already established parameter range in literature, validating our measurement approach.

\begin{table}
    \centering
    \begin{tabular}{lc}
        \hline
        Reference & SS branching ratio $r_S$\\
        \hline
        this work & 1.15 -- 2.26~\textsuperscript{$  $} \\
        Tetienne \etal{} \cite{tetienne12} & 0.0 -- 1.9~\textsuperscript{$\dagger$} \\
        Robledo \etal{} \cite{robledo11njp} & 1.1 -- 2.0~\textsuperscript{$  $} \\
        Gupta \etal{} \cite{gupta16} & 2.0 -- 2.3~\textsuperscript{$  $}\\
        Kalb \etal{} \cite{kalb18} & 2.0 -- 8.0~\textsuperscript{$\dagger$} \\
        Wirtitsch \etal{} \cite{wirtitsch23} & 13.3~\textsuperscript{$  $} \\
        \hline
    \end{tabular}
    \quad
    \begin{tabular}{lc}
        \hline
        Reference & ES coupling strength $\eta$ ($\si{\per\micro\second\per\cubic\milli\electronvolt}$)\\
        \hline
        this work & 176 -- 249 \\
        Plakhotnik \etal{} \cite{plakhotnik15} &  144\\
        Abtew \etal{} \cite{abtew11} & 149 \\
        Goldman \etal{} \cite{goldman15} & 276 \\
        \hline
            &    \\
            &   \\
    \end{tabular}
\caption{\textbf{Comparison to literature values} - Our calibrated values of $r_S$ (left) and $\eta$ (right) are compared to measurement values from other studies. For each study, we give the maximum parameter range (where applicable), considering measurement error and measurements on multiple NVs. Values denoted with $\dagger$ were transformed from a 7-level system convention to the 5-level system used in this work.} \label{tab:literaturecomp}
\end{table}

\subsection{Simulations and fits}
\label{sec:extrapol}
From the fitting process described above, we obtain all quantities needed to model the PL and contrast of each NV center in dependence of temperature, magnetic field, and strain. When performing these simulations, we fix the \textit{environment-sensitive} parameters to values obtained at base temperature. For \textit{PL vs. $B$}, these are the values given at the bottom of table \ref{tab:fitedparams}. For contrast simulations, we use the lowest-temperature values in Fig.~\ref{fig:SI-setup} at the respective magnetic field. Based on this set of parameters, we perform simulations of PL and contrast at all temperatures.

For \textit{PL vs. $B$}, this process results in the maps shown in Fig.~3(a) in the main text for \cI{} and Fig.~\ref{fig:SI-2Dmaps-E4-E2-G5}(a-c) for all other NVs. The maps emphasize the rich strain-dependence of the PL. We also note that when comparing the similarly-strained \cI{} and \eII{} (Fig.~\ref{fig:SI-2Dmaps-E4-E2-G5}(b)), we predict a feature in the form of a small peak at $\sim 50\unitformat{mT}$ and $\sim 50\unitformat{K}$ in the map of \eII{} which arises due to a different in-plane strain angle $ \phi_\delta $~\cite{ernst23modeling}.

We performed \textit{PL vs. $B$} measurements at select temperatures covering all three regimes (I-III) discussed in the main text, which we can compare with the theory by only refitting the \textit{environment-sensitive} parameters $b$ and $R$, while otherwise using results of the calibration. This has approximately the same effect as defining a scaling and an offset on our PL data and thus corrects for setup drifts.
We find good agreement between measurement and fits for all NVs (Fig.~3(b) of the main text and Fig.~\ref{fig:SI-2Dmaps-E4-E2-G5}(d-f)). We were, however, not able to resolve some small features such as the peak predicted for \eII{}, likely due to a combination of remanent field effects, drift in setup parameters during one sweep, and pulse-tube vibration noise limiting our resolution of PL.
We assumed the same magnetic field alignment for all measurements, i.e. we exclusively use the magnetic field orientation from the calibration fit. However, some drift and/or hysteresis of the vector magnet becomes apparent from the mismatch between data and fit in the width of the dips at the the ground state LAC ($\sim 100\unitformat{mT}$) in Fig.~\ref{fig:SI-2Dmaps-E4-E2-G5}(d). 

\begin{figure*}
    \includegraphics{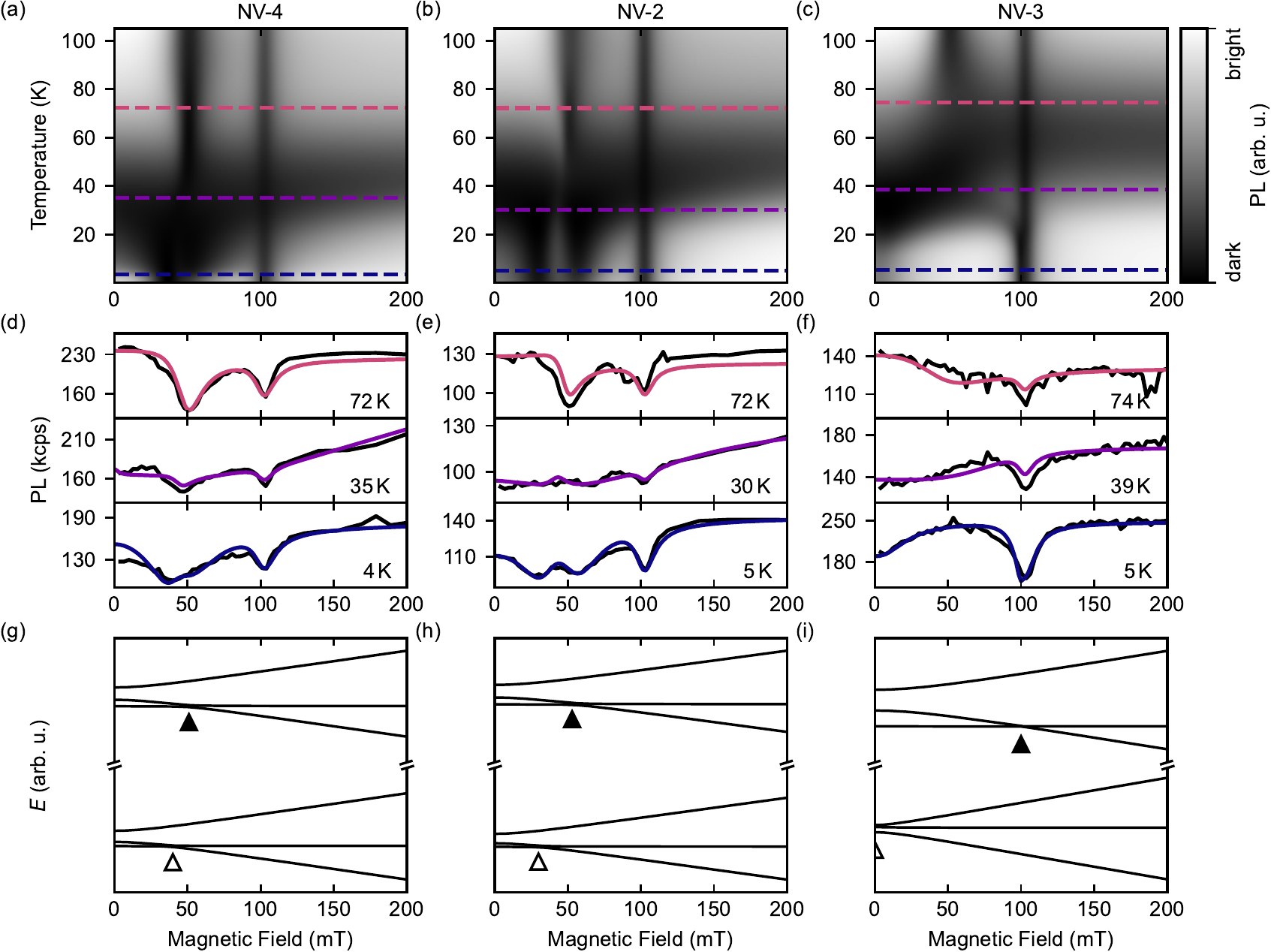}
    \caption{\textbf{Temperature dependence of the steady-state PL intensity} - Corresponding to Fig.~3 of the main text. Shown are simulations (a-c), measurements with fits (d-f) and level structures (g-i) for \eIV{}, \eII{}, and \gV{}, respectively.
		}
    \label{fig:SI-2Dmaps-E4-E2-G5}
\end{figure*}

To perform simulations of contrast versus temperature, we simulate and evaluate \textit{time-resolved pulsed ODMR} sequences at the desired temperatures. In the main text, we find a good match between our simulation and the measurement data, despite the fact that the simulation uses fixed \textit{environment-sensitive} parameters, whilst we find significant variation in the calibration at different temperatures (Fig.~\ref{fig:SI-setup}). This emphasizes the great stability of contrast measurements against setup drifts. 

Finally, for Fig.~1(a) of the main text we use the parameters of the calibration and only (re-)fit the \textit{environment-sensitive} parameters of the time-resolved pulsed ODMR traces shown in Fig.~1(a) together with their respective saturation measurement (with shared $ b $ and $ R $ and individual $ A $ as done in the calibration above).

\subsection{Influence of the phonon model on our simulation}
\label{sec:probing}

\begin{figure*}
    \includegraphics{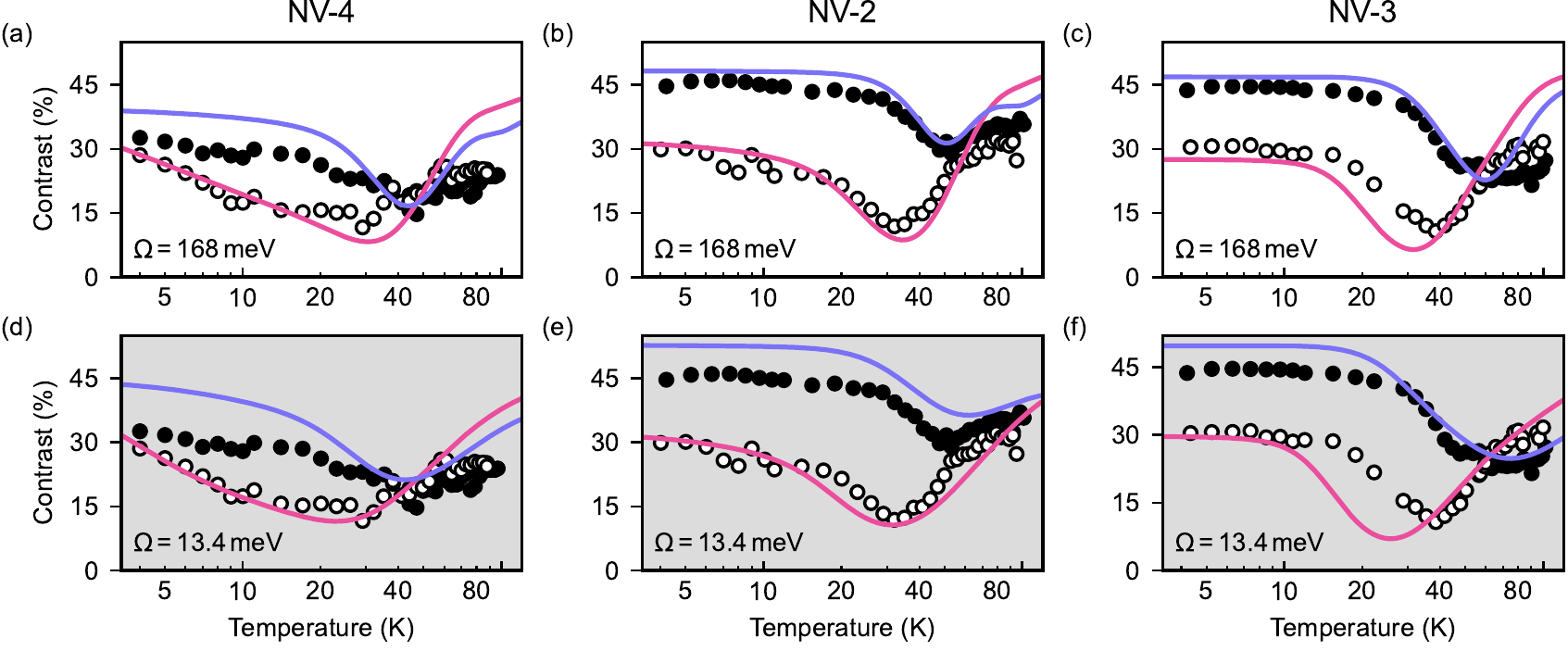}
    \caption{\textbf{Comparison between different cut-off energies} - Shown are simulations of temperature-dependent contrast (solid lines) of \eIV{}, \eII{}, and \gV{}, with two different phonon cut-off energies $\Omega$. For (a-c) $\Omega_\text{D} = 168\unitformat{meV}$ as in the main text, for (d-f) $\Omega_\text{P} = 13.4\unitformat{meV}$~\cite{plakhotnik15}. Empty (filled) circles are the same measurement data taken at $3 \unitformat{mT}$ ($200 \unitformat{mT}$) as presented in Fig.~3(e) of the main text.}
    \label{fig:SI-cutoff}
\end{figure*}

As mentioned in the main text, the temperature dependence of the orbital hopping rate is still under debate~\cite{gali19}. In Ref.~\cite{ernst23modeling} we give a detailed discussion of the usage of our model as a novel tool to investigate the nature of contributing phonon modes.
Here, we aim to put our findings in context with previous studies.

We have already found  good agreement of our fitted electron-phonon coupling strength $\eta$ with literature values (c.f. Table \ref{tab:literaturecomp}). However, we note that in all of these studies, different assumptions about the contributing phonon modes where made.
In particular, the integral over the phonon spectrum $I(T)$ for the two-phonon process (c.f. Eq.~2 of the main text or Eq.~\ref{equ:k2hoppUp} here) was in all cases solved in the Debye approximation but with a different phonon cut-off energy $\Omega$.
In this work, we assumed the Debye temperature of diamond ($\Omega_\text{D} = 168 \unitformat{meV}$) as the cut-off energy.
But in a study by \citet{plakhotnik15}, a much lower value of $\Omega_\text{P} = 13.4 \unitformat{meV}$ was found.

To investigate the effect of such a low cut-off energy, we use the value of $\Omega_\text{P}$ in our model and repeat the calibration of section~\ref{sec:fitting}.
As expected, it significantly alters the model, yielding higher fit results of $\eta$ (\eII{}: $ 284 \unitformat{\mu s}^{-1}\unitformat{meV}^{-3} $, \gV{}: $ 512 \unitformat{\mu s}^{-1}\unitformat{meV}^{-3} $, \eIV{}: $ 427 \unitformat{\mu s}^{-1}\unitformat{meV}^{-3} $) and a slower recovery towards room temperature.
Comparing the fits in Fig.~\ref{fig:SI-cutoff}, we find that using $\Omega_\text{D}$ yields better results at low temperatures and high magnetic fields, while -- due to the slower recovery -- $\Omega_\text{P}$ matches the high temperature data slightly better.
Thus, presumably due to the setup instabilities and uncertainty in temperature calibration at high temperatures in Setup A (see section \ref{sec:experimental}), the data presented in this work does not allow to definitely rule out either of the models. In the main text we exclusively present data fitted with $\Omega_\text{D}$, since the overall match appears better and the fitted coupling constant $\eta$ agrees well with literature values. We did not attempt to fit $\Omega$ in this work and expect that the correct phonon spectrum has to be used rather than a Debye model with fitted cutoff energy $\Omega$.
For future experiments, we expect a more stable contrast versus temperature measurement in the range $\SIrange{80}{150}{\kelvin}$ to give valuable insight into the contributing phonon modes.

\section{Additional simulations}

\subsection{Contrast, PL and sensitivty}
\label{sec:sensitivity}
In the main text we interchangeably used the PL and the contrast to observe or simulate the temperature dependent photo-physics of the NV center. In Fig.~\ref{fig:SI-2Dmaps}(a,b) we give a direct comparison of the temperature and magnetic field dependence of these two observables, demonstrating that they qualitatively show the same behavior.
Moreover, in Fig.~\ref{fig:SI-2Dmaps}(c) we show the corresponding sensitivity, a measure for the inverse signal-to-noise ratio (SNR) when using the NV center as sensor. In good approximation, it is inversely proportional to the contrast times the square root of the steady-state PL (Eq.~\ref{equ:sens}). Since both quantities are affected significantly by the spin relaxation process discussed in this work, the resulting sensitivity is impaired by more than a factor of 10.

\begin{figure*}
    \includegraphics{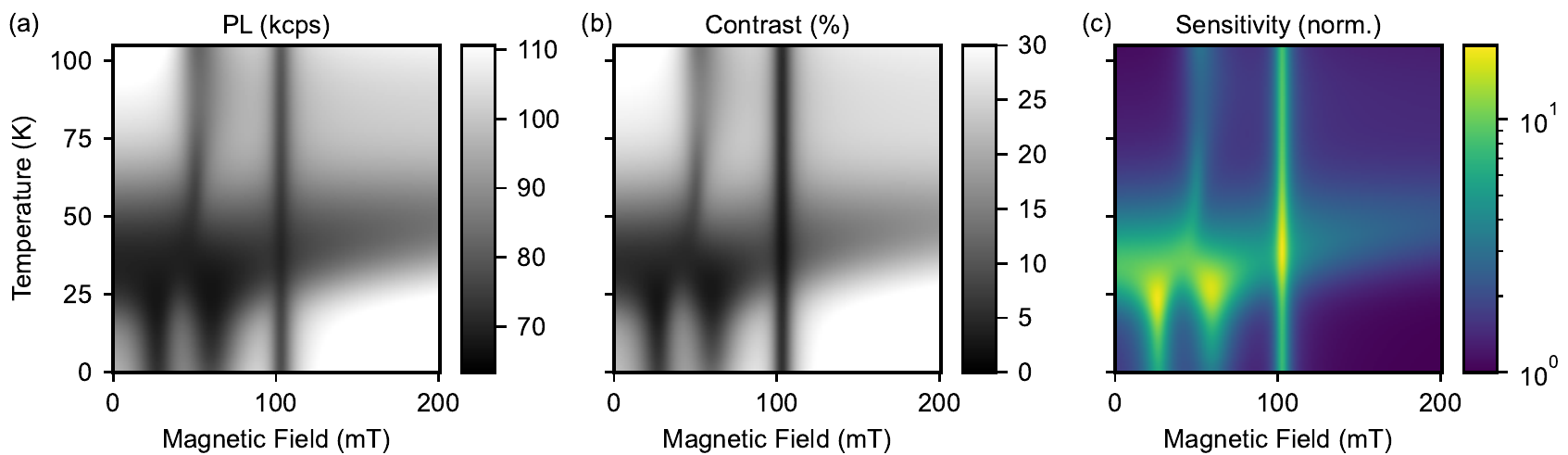}
    \caption{\textbf{Comparison of PL, contrast and sensitivity} - Shown are simulated maps of the steady-state PL (a) (same data as Fig.~3(c) of the main text), spin contrast $C$ (b) and normalized sensitivity $ s \propto \text{SNR}^{-1}$ (c) for parameters of \cI. All show qualitatively the same behavior. Since $ s \sim (C \sqrt{\text{PL}})^{-1}$, the strong variations in the NV centers sensitivity become apparent when combining (a) and (b), with its vast implications for measurements at intermediate temperatures.}
    \label{fig:SI-2Dmaps}
\end{figure*}

\subsection{Optimal integration time}
Since the NV center's photodynamics change with temperature, magnetic field, and strain, the integration time for optimal sensitivity in e.g. a pulsed ODMR measurement also changes. To reduce complexity, we chose to use the same integration time of $250\unitformat{ns}$ at all conditions. The effect of an integration time optimized for best sensitivity (for details see Eq.~\ref{equ:sens}) at the respective conditions has no influence on the qualitative behavior, as shown in Fig.~\ref{fig:SI-tint}.

\begin{figure*}
    \includegraphics{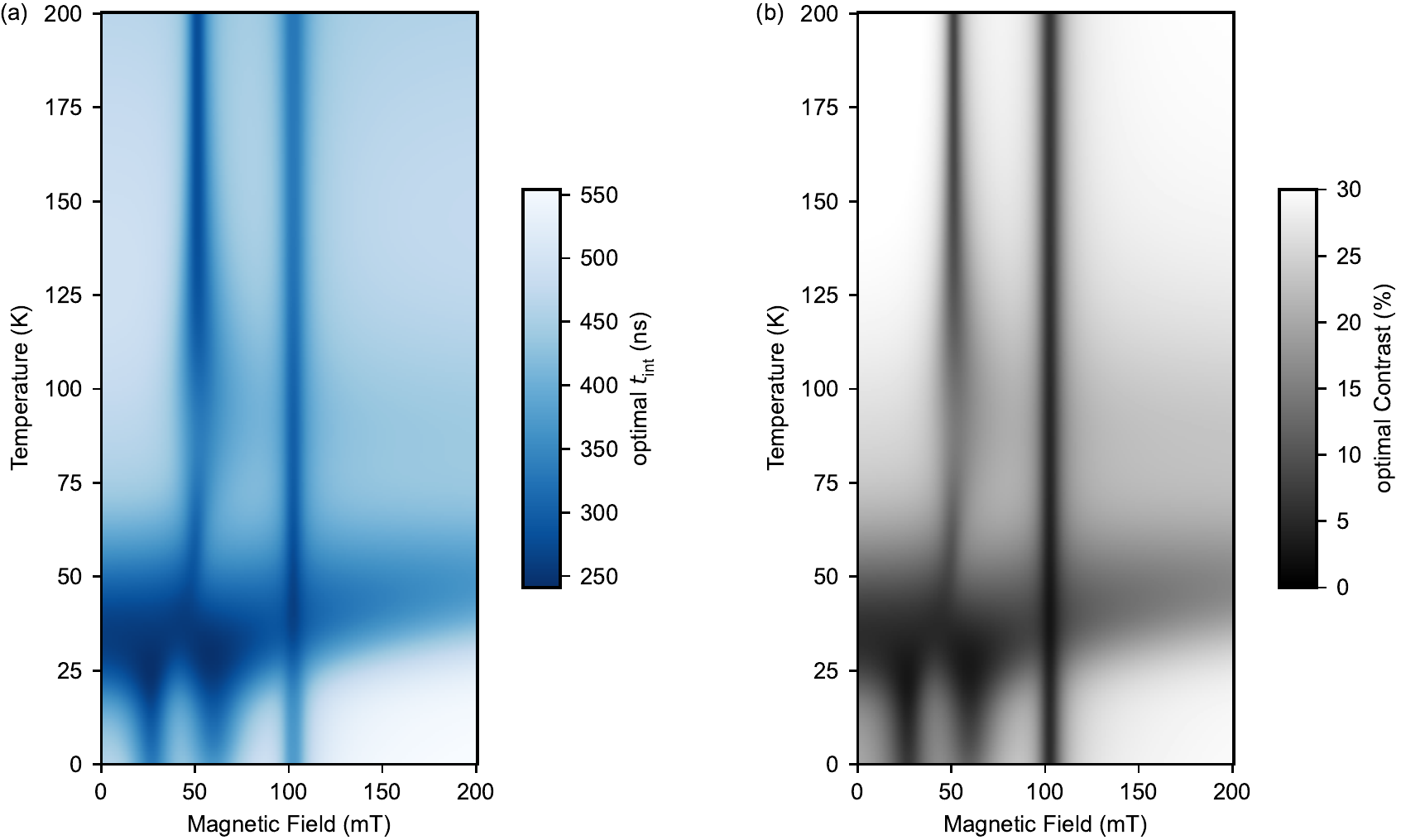}
    \caption{\textbf{Optimal integration time for \cI{}} - (a) At each field and temperature a simulated pulsed ODMR was optimized for best sensitivity (Eq.~\ref{equ:sens}) and the resulting integration time is plotted. (b) Corresponding spin contrast map.
    Both maps shows the same pattern as they are dominated by the same spin relaxation process addressed in this work. Spin relaxation in the ES leads to a faster loss of ODMR spin contrast under laser illumination, promoting a shorter integration time.
    For comparison, the same contrast map for $t_\text{int} = 250\unitformat{ns}$ is given in Fig.~\ref{fig:SI-2Dmaps}(b). While, as expected, a change of $t_\text{int}$ changes the absolute contrast value, the qualitative behavior is unaffected.} 
    \label{fig:SI-tint}
\end{figure*}

\subsection{Spin initialization and readout fidelity}
The temperature dependent spin relaxation process discussed in this work affects both, the spin state initialization (by laser illumination) as well as the spin state readout (as e.g. in pulsed ODMRs presented in Fig.~1(a) of the main text).
The spin contrast as presented in this work suffers from a combination of both. To illustrate the effect of the spin relaxation process on the initialization and readout individually, we plot them separately in Fig.~\ref{fig:SI-2DmapsInitReadout}.

\begin{figure*}
    \includegraphics{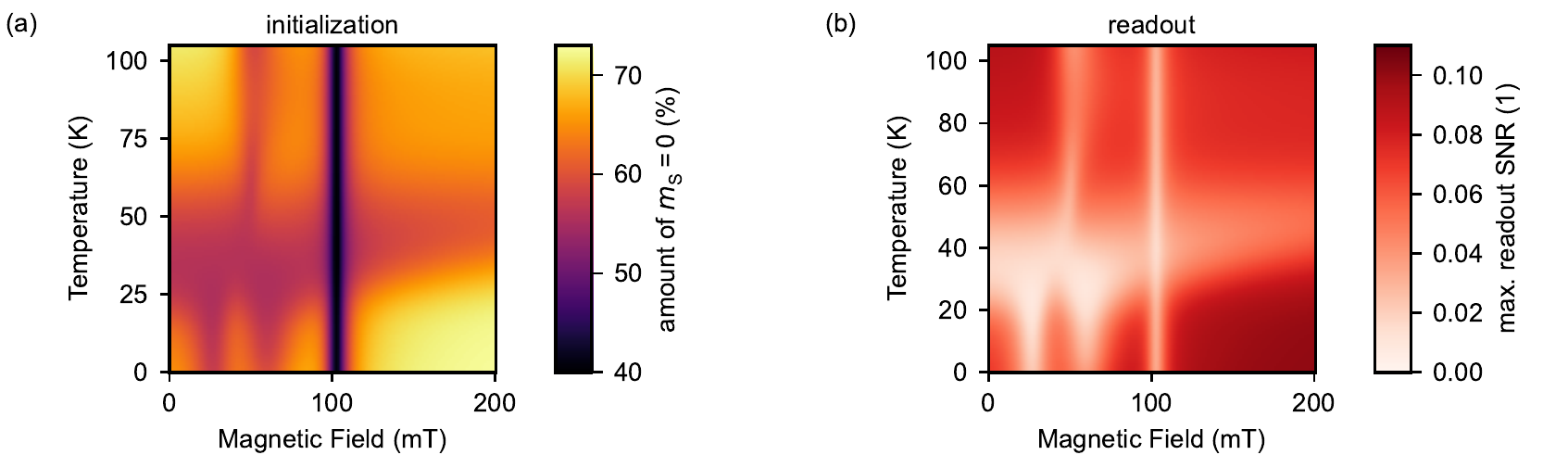}
    \caption{\textbf{Effect on spin state initialization and readout} - (a) Simulated map of the amount of \ms{0} after a long green laser pulse followed by a long waiting time (no spin state $ T_1 $ process is present). This is a common scheme to initialize the spin state of the NV center.
    (b) Simulated maximal signal-to-noise ratio of a single spin state readout~\cite{ernst23modeling}. 
    For this simulation we assumed a perfect initialization with an amount of \ms{0} of 1. The integration time $ t_\text{int} $ was optimized at each point for maximal signal-to-noise ratio of the readout, which is almost identical to optimizing it for best sensitivity as done in Fig.~\ref{fig:SI-tint}(a) (c.f. Eq.~\ref{equ:sensIdeal}).
    The combination of both (a) and (b) give rise to the reduced performance presented in Fig.~\ref{fig:SI-2Dmaps}.
    Parameters of \cI{} were used.}
    \label{fig:SI-2DmapsInitReadout}
\end{figure*}

\subsection{Extrapolation to low strain or high magnetic field}
\label{sec:strain}
As discussed in the main text, very high magnetic fields are required to completely suppress the detrimental effect of spin mixing at intermediate temperature. This can be seen in Fig.~\ref{fig:LowStrainAndHighBMaps}(a) on a map of the NV center sensitivity up to $ \SI{5}{\tesla} $.
Further, the main text claims that the dip in performance at intermediate temperature is common to all NV centers. To that end, we plot in Fig.~\ref{fig:LowStrainAndHighBMaps}(b) the sensitivity for an NV center that has very low strain, as found in bulk diamond samples.
We plot sensitivity instead of $\text{PL}$ or spin contrast here, as it is proportional to the inverse SNR and therefore the relevant quantity when comparing the performance of the \ch{NV-} center under different conditions. In section \ref{sec:sens}, a brief derivation of the sensitivity relation (Eq.~\ref{equ:sens}) used here is provided.

\begin{figure}
    \centering
    \includegraphics{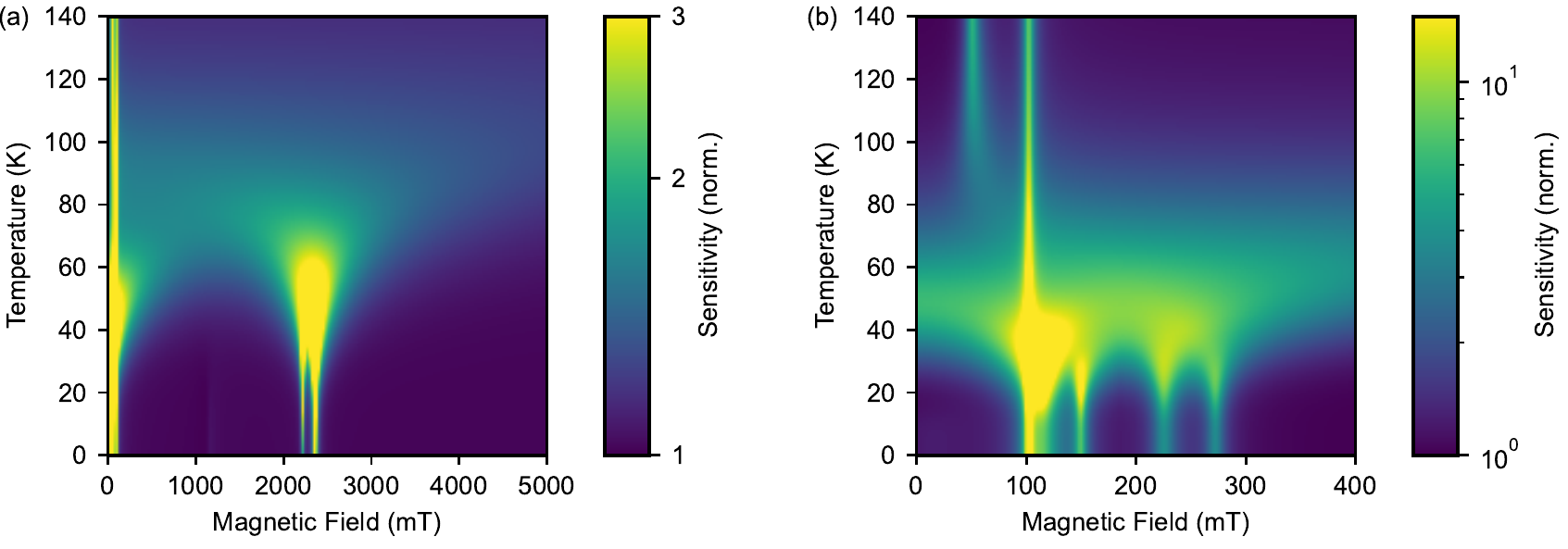}
    \caption{\textbf{Model predictions for extreme cases} - (a) Normalized sensitivity map for parameters as found for \cI{} extended to very high magnetic field. Only when approaching $ \sim \SI{5}{\tesla} $, the dip in performance at intermediate temperatures gradually disappears. We note though, that this field value is significantly influenced by the position of the LACs above $ \SI{2}{\tesla} $, which where observed to be shifted to significantly higher fields supposedly due to a strain dependent orbital $ g_l $ factor~\cite{happacher22}.
    (b) Normalized sensitivity map for a NV center with very low strain splitting of $ \SI{1}{\giga\hertz} $ ($ \delta_\perp = \SI{0.5}{\giga\hertz} $) as found in bulk diamond. Apart from the strain, parameters are as found for \cI{}. The same qualitative behavior is observed as discussed in the main text, though the dip in performance is shifted to higher temperature. For both plots, the integration time was optimized for best sensitivity at each field and temperature for a comparison of performance.
    }
\label{fig:LowStrainAndHighBMaps}
\end{figure}

\section{Spin mixing amplitudes}

In the main text, we introduce the spin mixing amplitude $\epsilon_{\ket{i},\ket{j}}$ to describe the superposition between two basis states, where one of them contributes only a small amount. In a Bloch sphere picture, this amounts to only a small inclination from one of the poles (where the poles are given by the two basis states)~\cite{ernst23modeling}. In this picture, the new eigenstate $\ket{\text{es}}$ is given by
\begin{equation}
    \ket{\text{es}} = \ket{i} + \epsilon_{\ket{i}, \ket{j}}\ket{j}\,.
\end{equation}
The new state $\ket{\text{es}}$ is not properly normalized, but for sufficiently small  $\epsilon_{\ket{i},\ket{j}}$, this is a good approximation of the actual eigenstate $\ket{\tilde{\text{es}}}$. We calculate the  $\epsilon_{\ket{i},\ket{j}}$ by projecting the basis states onto the various eigenstates, such that (for a given eigenstate)
\begin{equation}
    \epsilon_{\ket{i},\ket{j}} = \frac{\braket{j}{\tilde{\text{es}}}}{\braket{i}{\tilde{\text{es}}}}\,.
\end{equation}
In Tab.~\ref{tab:epsilonij}, we give the absolute squares of all relevant $\epsilon_{\ket{i},\ket{j}}$, given the setting of Fig.~2(c).

\begin{table}[]
    \centering
    \begin{tabular*}{0.45\textwidth}{@{\extracolsep{\fill}}|l|ccccc|}
        \hline
         $\left|\epsilon_{\ket{i},\ket{j}}\right|^2$  & $\ket{\x}\ket{+}$ & $\ket{\x}\ket{0}$ & $\ket{\y}\ket{-}$ & $\ket{\y}\ket{+}$ & $\ket{\y}\ket{0}$ \\
         \hline
         $\ket{\x}\ket{-}$ &    0.     & 0.003     & 0.    & 0.006 & 0.  \\
         $\ket{\x}\ket{+}$ &      &    0.      & 0.003 & 0.     & 0. \\
         $\ket{\x}\ket{0}$ &        &           &  0.     &  0.     & 0. \\
         $\ket{\y}\ket{-}$ &        &           &       &    0.       & 0.120\\
         $\ket{\y}\ket{+}$ &        &           &       &          & 0.\\
         \hline
    \end{tabular*}
    \caption{Absolute squares of the spin mixing amplitudes $\left|\epsilon_{\ket{i},\ket{j}}\right|^2$ for the setting of Fig.~2(c) in the main text. Only the $\ket{\y}\ket{0}$ and $\ket{\y}\ket{-}$ states mix significantly.}
    \label{tab:epsilonij}
\end{table}

\section{Strain dependence of orbital hopping rates}
\label{sec:hoppingPlots}
\begin{figure}
	\centering
	\includegraphics{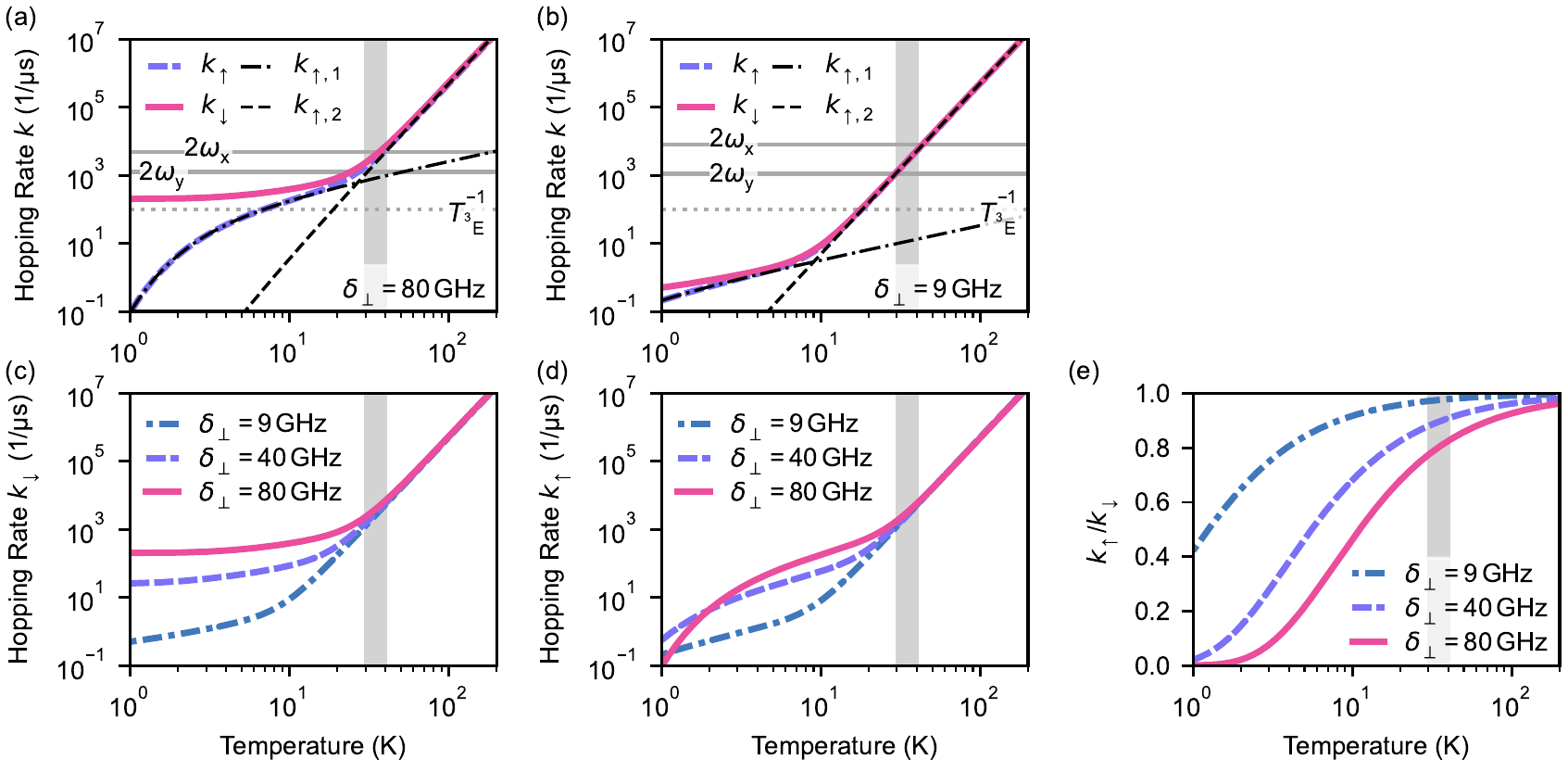}
	\caption{\textbf{Orbital hopping rates under different conditions} - 
	(a) Rates up and down and contributions to the up rate of the one- and two-phonon processes for high strain. Also shown are the inverse optical lifetime $T_{\mathrm{^3E}}^{-1}$ and twice the Larmor frequencies $2\omega_{x(y)}$ (in MHz) as horizontal lines. Unless specified otherwise, the same setting as in Fig.~2(c) and (d) of the main text is used in the entire figure. The $\SIrange[range-phrase=-]{30}{40}{\kelvin}$ region where highest spin relaxation is observed is marked (gray shading). 
    (b) Same for low strain. In both (a) and (b), the same overlap of the region of highest spin relaxation with the range where twice the Larmor frequencies and rates match is observed as explained for medium strain in the main text.
    (c) Effect of strain on the hopping rate down.
    (d) Effect of strain on the hopping rate up.
    (e) Ratio of rate up and down for different strains.
	}
	\label{fig:SI-hopping-rates}
\end{figure}

As the orbital hopping rates -- paired with the spin mixing discussed in the main text -- cause the observed spin relaxation process, we give a broader overview of their strain dependence here.
The temperature dependence of the orbital hopping rates up ($k_\uparrow$) and down ($k_\downarrow$) are plotted in Fig.~\ref{fig:SI-hopping-rates}. We use the same parameter set as in Fig.~2(d) of the main text, but vary the in-plane strain $ \delta_\perp $. 
In Fig.~\ref{fig:SI-hopping-rates}(a), we present a similar setting as in Fig.~2(d), but for high strain $\delta_\perp = \SI{80}{\giga\hertz}$. In Fig.~\ref{fig:SI-hopping-rates}(b) the same setting is also plotted for our low strain case of $\delta_\perp = \SI{9}{\giga\hertz}$.

At high strain in Fig.~\ref{fig:SI-hopping-rates}(a), as discussed in the main text, two effects can be observed.
First, at low temperature we find a significantly higher $k_\downarrow$ for high strain $\delta_\perp$ (Fig.~\ref{fig:SI-hopping-rates}(c)).
This is caused by the increase of the phonon mode density with energy.
If the orbital branch splitting $\sim 2 \delta_\perp$ is larger, a higher mode density is available for the spontaneous one-phonon emission process (c.f. second term in Eq.~1 of the main text).
Second, we also find a rapid increase of $k_\uparrow$ at high strain with rising temperature (Fig.~\ref{fig:SI-hopping-rates}(d)).
At very low temperature and high strain, the phonon modes required for the one-phonon absorption process are not populated.
But with rising temperature, they rapidly get thermally activated and $k_\uparrow$ reaches the high value of $k_\downarrow$.
This can be seen in the detailed balance of the rates in Fig.~\ref{fig:SI-hopping-rates}(e), yielding a Boltzmann distribution of population.
This combination of, first, a high $k_\downarrow$ with, second, a rapidly increasing $k_\uparrow$ below $\SI{10}{\kelvin}$ gives rise to the observed drop of spin contrast in Fig.~3(e) of the main text.

\end{widetext}


\end{document}